\documentclass[preprint,12pt]{elsarticle}

\usepackage{physics}
\usepackage{graphicx}
\usepackage{caption}
\usepackage{subcaption}
\usepackage{amsmath}
\usepackage{latexsym}
\usepackage{mathalpha}
\usepackage[colorlinks=true,linkcolor=blue]{hyperref}
\hypersetup{pdfauthor={Name}}
\usepackage{multirow}
\usepackage{array}
\usepackage{float}
\usepackage{placeins}
\newcommand{\PreserveBackslash}[1]{\let\temp=\\#1\let\\=\temp}
\newcolumntype{C}[1]{>{\PreserveBackslash\centering}p{#1}}

\usepackage{amssymb}
\usepackage{MnSymbol,bbding,pifont}

\usepackage{lineno}
\usepackage[T1]{fontenc}
\biboptions{sort&compress}

\journal{---}

\begin{document}

\begin{frontmatter}

\title{How ``mixing'' affects propagation and structure of intensely turbulent, lean, hydrogen-air premixed flames}


\author[fir]{Yuvraj}
\author[sec]{Hong G. Im}
\author[fir]{Swetaprovo Chaudhuri \corref{cor4}}
\ead{swetaprovo.chaudhuri@utoronto.ca}

\address[fir]{Institute for Aerospace Studies, University of Toronto, Toronto, Canada}
\address[sec]{Clean Combustion Research Center, King Abdullah University of Science and Technology, Thuwal, Saudi Arabia}

\cortext[cor4]{Corresponding author}

\begin{abstract}
Understanding how intrinsically fast hydrogen-air premixed flames can be rendered much faster in turbulence is essential for the systematic development of hydrogen-based gas turbines and spark ignition engines. Here, we present fundamental insights into the variation of flame displacement speeds by investigating how the disrupted flame structure affects speed and vice-versa. Three DNS cases of lean hydrogen-air mixtures with effective Lewis numbers ($Le$) ranging from about 0.5 to 1, over Karlovitz number ($Ka$) range of 100 to 1000 are analyzed. Suitable comparisons are made with the closest canonical laminar flame configurations at identical mixture conditions and their appropriateness and limitations in expounding turbulent flame properties are elucidated. Since near zero-curvature surface locations are most probable and representative of the average flame geometry in such large $Ka$ flames, statistical variation of the flame displacement speed and concomitant change in flame structure at those locations constitute the focus of this study. To that end, relevant flame properties are averaged in the direction normal to the zero-curvature isotherm locations to obtain the corresponding conditionally averaged flame structures. In the leanest case with smallest $Le$, the temperature increases beyond that of the standard laminar flame downstream of the zero-curvature regions, leading to enhanced local thermal gradient and flame speeds in the conditionally averaged structure. These result from increased heat-release rate contribution by differential diffusion ($Le \ll 1)$ in positive curvatures downstream of the zero-curvature locations. Furthermore, locally, the flame structure is broadened for all cases due to a reversal in the direction of the flame speed gradient. This reversal is caused by cylindrical flame-flame interactions upstream of the zero-curvature regions, resulting in localized scalar mixing within the flame structure. The combined effect of these two non-local phenomena defines the conditionally averaged flame structure and the associated variation of the local flame speed of a premixed flame in turbulence.

\end{abstract}


\end{frontmatter}
\newpage
\section*{Novelty and Significance Statement}      
\noindent

The paper presents fundamental discoveries pertaining to the structure and propagation of intensely turbulent, lean premixed hydrogen-air flames, emerging from the analysis of averaged flame structures conditioned to zero-curvature surface locations. These locations are most probable alongside corresponding to the mean of curvature distribution. Analysis of such structures reveals how non-local effects determine the average flame displacement speed, for the first time. Non-local effects appear in fluid turbulence literature but rarely in turbulent combustion. The paper shows that non-local effects within the flame structure address long-standing questions on how premixed flames are broadened in turbulence and why local flame displacement speeds of ultra-lean hydrogen-air flames are ubiquitously higher than their standard laminar counterpart.

\section*{Author Contributions}

\begin{itemize}

  \item{Y performed research, analyzed data, wrote the paper}

  \item{HGI performed research, wrote the paper }

  \item{SC designed research, performed research, wrote the paper}

\end{itemize}

\newpage

\section{Introduction}
\label{S:1}
Hydrogen is considered a promising alternative fuel to achieve sustainable energy solution for both transportation and stationary applications. Ultra-lean premixed combustion of hydrogen has the potential of achieving high efficiency and low NO$_x$ emissions \cite{gkantonas2023hydrogen}. Still, some critical challenges need to be overcome before its full implementation in practical engineering systems. One notable issue is the extremely high flame speed that often occurs erratically, in the presence of strong turbulence, leading to flashback and damage to the gas turbine injector system. A similar situation is encountered for SI engines where high flame speed and reactivity of stratified hydrogen-air mixtures may result in pre-ignition or knocking. 

Earlier studies on local flame displacement speeds investigated the effect of local curvature \cite{echekki1996, echekki1999, im2000effects, chen1998correlation, cifuentes2018, chakraborty2005_1, uranakara2016, uranakara2017} and strain rate\cite{echekki1996, chen1998correlation, peters1998statistics}, Lewis number \cite{chakraborty2005influence, haworth1992, rutland1993,alqallaf2019}, and thickness \cite{sankaran2015response} in the context of turbulent flames. Using direct numerical simulations (DNS) of lean H$_2$-air premixed turbulent flames for $\phi=0.3-0.4$ \citet{aspden2011_2} investigated the variation of local and global burning rates over a large $Ka$ range: $10 \leq Ka \leq 1562$ with focus on differential diffusion and transition to distributed burning. Consumption speeds of ultra-lean hydrogen-air turbulent premixed flames were comprehensively studied \cite{amato2015topology} and compared with canonical laminar flame configurations, focusing on positive curvatures. Furthermore, the importance of zero-curvature regions was emphasized and the departure of averaged local consumption speed behavior of high $Ka$ turbulent flames from that of the strained laminar flames was highlighted \cite{Amatophdthesis}. \citet{day2009} investigated lean H$_2$-air flames at an equivalence ratio of $\phi = 0.37$ in two different conditions: moderate turbulence and in the absence of turbulence, where the intrinsic thermo-diffusive instability governed the dynamics. The study found that the most probable consumption speeds at zero curvature were significantly enhanced over the standard laminar flame speed for both conditions. A recent study for lean hydrogen-air flames \cite{howarth2023thermodiffusively} reported that both freely propagating and turbulent flames of lean hydrogen-air mixtures demonstrated similar thermo-diffusive responses in moderate $Ka$ regimes. Higher local equivalence ratios and higher consumption speeds characterized the positively curved regions on the flame surface. It was also argued that enhanced temperatures and consumption speed in zero-curvature regions occurred due to higher heat release rate in the upstream, preceding positive curvature regions.

Recent Lagrangian analysis tracked the evolution of flame surface points or flame particles \cite{chaudhuri2015life} from positively curved leading regions \cite{zeldovich1985mathematical, amato2014analysis, dave2018} to annihilation at negatively curved trailing regions \cite{chaudhuri2015life, dave2020}. It was shown that for moderate $Ka$, near-unity $Le$ flames, flame-flame interaction leading to annihilation at large negative local curvatures $\kappa$ results in very large excursions of the density-weighted flame displacement speed $\widetilde{S_d}=\rho_0 S_d/\rho_u$ over its unstretched, standard laminar value $S_L$ \cite{dave2020}. Here, $\rho_u$ and $\rho_0$ are the density of the reactant mixture and the density at the isotherm of interest, $T=T_0$, within the flame structure. Once such local self-interaction begins, the enhanced $\widetilde{S_d}$ leads to larger negative $\kappa$, resulting in further enhanced $\widetilde{S_d}$. This fast-paced process continues until the flame surface is annihilated. The local flame structure differs greatly from an unstretched standard laminar flame during the interactions. Given the distinct transient structure, the steady weak stretch theory \cite{bechtold2001, giannakopoulos2015_1, giannakopoulos2015_2} falls short in explaining the aforementioned enhanced $\widetilde{S_d}$ at large negative curvatures. As such, an interacting flame theory was proposed to model the $\widetilde{S_d}$ at large negative curvatures ~\cite{dave2020, yuvraj2023} as:

\begin{equation}
\widetilde{S_d}=-2\widetilde{\alpha_0}\kappa
\end{equation}\label{Eq_Sd_I}

\noindent where curvature, $\kappa=\kappa_1+\kappa_2$ and $\kappa_1$ and $\kappa_2$ are the minimum and maximum principal curvatures respectively. $\alpha_0$ is the thermal diffusivity defined at $T_0$, the isotherm of interest. $\widetilde{\alpha_0}$ is the corresponding density-weighted thermal diffusivity.
The proposed model Eq.~(\ref{Eq_Sd_I}), a linear function of the local curvature, could capture the behavior of $\widetilde{S_d}$ at large negative curvatures $\kappa\ll0$. Linear scaling with curvature had been previously proposed \cite{Peters2000, chakraborty2009direct} without the pre-factor 2. 
For any point on an isotherm, local flame-displacement speed $S_d$ can be defined based on the right-hand side of the energy equation (Eq.~(\ref{Eq_2})) as:

\begin{equation}
\begin{split}
    S_d &= \frac{1}{\rho C_p}\left[ \frac{\grad\cdot(\lambda^\prime\grad T)}{|\grad T|} \right.
     -\frac{\rho \grad T \cdot \sum_{k}^{}  ( C_{p,k} V_k Y_k)}{|\grad T|}
     \left.-\frac{\sum_{k}^{} h_k \dot{\omega}_k}{|\grad T|} \right]
    \end{split}
    \label{Eq_2}
\end{equation}

\noindent where $\lambda^\prime$, $V_k$, $h_k$ and $\dot{\omega}_k$ are the thermal conductivity of the mixture, the molecular diffusion velocity, enthalpy and net production rate of the $k$th species at a given isotherm, respectively. $C_{p,k}$ and $C_{p}$ are the constant-pressure specific heat for the $k$th species and for the bulk mixture, respectively. 
For an isotherm lying in the preheat zone, $S_d$ can be approximated as:

\begin{equation}\label{Eq_Sd_I_Diff}
S_d\approx \frac{1}{\rho C_p}\Big[ \frac{\grad\cdot(\lambda^\prime\grad T)}{|\grad T|}\Big]
\end{equation}

\noindent Eq.~(\ref{Eq_Sd_I_Diff}) can be further decomposed into curvature and gradient terms following \cite{echekki1999} as:

\begin{equation}\label{Eq_Sd_I_Diff_2}
S_d\approx -\alpha\kappa -\frac{\alpha}{|\grad T|} \boldsymbol{n}\cdot \grad (|\grad T|)
\end{equation}

\noindent where $\kappa=-1/r$, $r$ being the local radius of curvature. For a cylindrical, inwardly propagating flame $\boldsymbol{n}\cdot \grad = \partial/\partial n = -\partial/\partial r$ and $|\grad T|=-\partial T/\partial n= \partial T/\partial r$, where $\boldsymbol{n}=-\grad T/|\grad T|$ is the local normal defined positive in the direction of the unburnt mixture. During local flame-flame interaction as $r\rightarrow 0$, since $T$ is locally radially symmetric, we have:

\begin{equation}
    \lim_{r\to0} |\grad T| = \lim_{r\to0} \frac{\partial T}{\partial r} =0 
\end{equation}

\noindent Using Bernoulli-L'H\^opital rule we get:

\begin{equation}\label{Eq_second_derivative_limit}
  \lim_{r\to0} \frac{\partial T/\partial r}{r} = \lim_{r\to0} \frac{\partial^2 T}{\partial r^2}
\end{equation}

\noindent Therefore, during flame-flame interaction as $r \rightarrow0$ the second term in Eq.~(\ref{Eq_Sd_I_Diff_2}) can be simplified using Eq.~(\ref{Eq_second_derivative_limit})
\cite{buenzli2022} as: 

\begin{equation}
-\frac{\alpha}{|\grad T|} \boldsymbol{n}\cdot \grad (|\grad T|) \sim  \alpha \frac{\partial^2 T/\partial r^2}{\partial T/\partial r} \sim \alpha \frac{(\partial T/\partial r)/r}{\partial T/\partial r} \sim \frac{\alpha}{r} \sim -\alpha\kappa
\end{equation}\label{Eq_diffusion_3}

\noindent Thus, scaling arguments show that the curvature and the normal gradient term contribute equally as  $r\rightarrow0$ or $\kappa \rightarrow -\infty$; where we recover $\widetilde{S_d}=-2\widetilde{\alpha_0}\kappa$. This is Eq.~\ref{Eq_Sd_I} obtained for $\kappa \delta_L \ll 0$. Flame-flame interactions at large negative curvatures are an important phenomenon in their own right since, typically, the large heat-release fluctuation resulting from rapid flame annihilation leads to sound generation \cite{haghiri2018sound,brouzet2019annihilation}. 

Flame surfaces undergoing self-interactions have been shown to form a variety of local topologies \cite{dunstan2013,griffiths2015, trivedi2019topology, brouzet2019annihilation,haghiri2020}. Extremely large negative and positive curvatures lead to tunnel closure and tunnel formation, respectively. \citet{haghiri2020} found enhanced $S_d$ during island burnout, pinch-off and tunnel formation. When the $Ka$ is small, such interactions are infrequent and limited to rare localized events. However, increasing $Ka$ makes such interactions frequently encountered among individual iso-scalar surfaces. As such, at $Ka \sim\mathcal{O} (1000)$, localized flame-flame interaction characterized by $\kappa \delta_L \leq -1$ was prevalent in about 44\% of the flame structure~\cite{yuvraj2023}. 

Considering its significance, a theoretical analysis of the interacting preheat zones of cylindrical, laminar, premixed, inwardly propagating flame (IPF) extended and quantified \cite{dave2020} the curvature-based scaling of flame displacement speed \cite{Peters2000}. As mentioned above, at very large $Ka$, the self-interactions become highly frequent and localized at the iso-surface level~\cite{yuvraj2023}, but the curvature-based scaling was found to remain valid statistically \cite{yuvraj2022local}. In view of this, the surprising finding that a simple 1D laminar flame model could successfully describe the statistically averaged flame speed characteristics at large negative $\kappa$ in turbulence may be justified by several factors. First, flame-flame interaction is a fast event during which the local turbulence remains frozen. Furthermore, at substantial curvature values, the local radius of curvature assumes length scales comparable to the Kolmogorov scales. This also makes DNS an appropriate choice as it resolves all the scales of turbulence. Finally, turbulence folds the propagating flame surfaces along lines into a locally cylindrical form \cite{pope1989curvature, zheng2017principal, yuvraj2023}; hence large curvature surfaces in turbulence are typically cylindrical. Thus an IPF configuration serves as a canonical model to describe the statistical behavior of $\langle \widetilde{S_d}|\kappa \rangle$ for large negative curvatures undergoing flame-flame interaction. 

Most importantly, such flame-flame interaction at the entire flame structure level or the internal isotherm level leads to vanishing scalar gradients. This could be considered an internal \emph{mixing} process, resulting in the broadening of the internal flame structure. Broadening of the flame structure or the separation between the iso-surfaces within the structure is proportional to $1/|\grad c|$, where $\grad c$ is the gradient of the scalar $c$~\cite{hamlington2011, hamlington2011b}. Previous works \cite{hamlington2011, hamlington2010, poludnenko2010, poludnenko2011, song2020dns, lipatnikov2021} reported substantial broadening in the preheat zone with a comparatively low broadening in the reaction zone, even under extreme $Ka$ conditions extending to the distributed regime. This was shown by a high probability of lower valued scalar gradient generated by strong mixing in the preheat zone when compared to the reaction zone. A comprehensive review of flame structure and scalar gradients for highly turbulent premixed flames have been presented in \cite{steinberg2021}. While the flame was broadened overall, regions with scalar gradients larger than the corresponding laminar counterpart were also found on the flame surface. \citet{su2003} experimentally investigated two-dimensional and three-dimensional scalar dissipation rate $\chi$ fields for planar propane jet. It was shown that the negative skewness of the $\chi$ probability density function is its intrinsic property rather than an artifact of the experiment. \citet{chaudhuri2017} reported the flame being thinned on average from reactants to products for lean H$_2$-air slot jet flame. This contrasting observation was attributed to the configuration of the flame, i.e., the effect of strong shear turbulence.   

In the present work, we unravel how mixing associated with self-interacting flame surfaces at large negative curvatures, as well as the differential diffusion at positive curvatures end up determining the local flame structure and speed at zero-curvature. The analysis is based on three direct numerical simulation (DNS) datasets at different Lewis and Karlovitz number conditions. The present flames are apparently free from the periodic cellular structures resulting from diffusive-thermal instability due to structural disruption by turbulence at very large $Ka$ \cite{lipatnikov2005molecular,chaudhuri2011,chomiak2023simple}, while recognizing that thermo-diffusive effects synergistically works with turbulence to affect flame structure and propagation \cite{berger2022synergistic}. Recent literature has comprehensively explored diffusive-thermal instability of hydrogen flames \cite{berger2022synergistic, rieth2023effect, song2020dns, howarth2023thermodiffusively, howarth2022empirical} and has shown how differential diffusion of hydrogen and associated radicals increase consumption and displacement speeds.

The paper is structured as follows. We begin by introducing the DNS configuration, numerical methods and the detailed methodology used to simulate the premixed flames datasets under investigation in Section \ref{sec: 2}. From the results in Section~\ref{sec: 3}, we observe a strong enhancement in $\widetilde{S_d}$ in the regions with large negative curvatures as well as with zero-curvatures for the ultra-lean case. To understand the enhancement of $\widetilde{S_d}$, we compare the DNS results with their respective canonical laminar configurations based on the local curvature in the following subsections. These comparisons unravel important insights, particularly the limitations of one of the canonical models in explaining the DNS results in zero-curvature regions. Further analysis suggests that the difference lies in the flame structure. To that end, we explore the conditionally averaged flame structure in the neighborhood of the zero curvature isotherms along the local normal directions which leads to the following crucial findings. In the zero-curvature regions, $\widetilde{S_d}$ is enhanced due to increased heat release rate in the positively curved downstream locations within the conditionally averaged flame structure. Furthermore, such enhancement is also accompanied by local broadening due to mixing effected by non-localized flame-flame interactions in the upstream negatively curved locations within the flame. Section~\ref{sec: 4} concludes the paper.


\section{Methodology}
\label{sec: 2}

\begin{table}[h!]
\footnotesize
 \begin{center}
 \begin{tabular}{ | p{6cm} | p{1.8cm} | p{2cm} | p{1.6cm} | } 
 \hline
  Parameters & Le05Ka100 & Le08Ka1000 & Le1Ka100\\  \hline
  \hline
  $T_u$ [K] & 310 & 300 & 310 \\
 $\phi$ & 0.4 & 0.7 & 0.81 \\
 $Le$ \cite{matalon_2003} & 0.48 & 0.76 & 0.93 \\
 Domain dimensions [cm] & $2.40$ & 0.43 & 1.12 \\
 & $\times0.60$ & $\times0.15$ & $\times0.28$\\
  & $\times0.60$ & $\times0.15$ & $\times0.28$\\
 Grid points & $800$ & $1645$ & $1120$ \\
  & $\times200$ & $\times560$ & $\times280$ \\
  & $\times200$ & $\times560$ & $\times280$ \\
 Integral length scale \cite{pope_turbulentflows}, $L_{11}$ [cm] & 0.3 & 0.03 & 0.11 \\
 Root mean square velocity, $u_{rms}$ [cm/s] & 507.7 & 4746.7 & 1893.9 \\
 Kolmogorov length scale, $\eta$ [$\mu$m] & 19.9 & 2.141 & 6.5 \\
 Karlovitz no., $Ka$ &  115.4 &  1125.7 & 100.1 \\
 Reynolds no., $Re_t$& 799.6 & 699.5 & 996.2 \\
 Damk\" ohler no., $Da$ & 0.25 & 0.02 & 0.32 \\
 $S_L$ [cm/s] & 25.23 & 135.62 & 183.82 \\
 $\delta_L$ [cm] & 5.98E-02 & 3.54E-02 & 3.53E-02 \\
 Cutoff wavelength$^*$, $\lambda_c$ [cm]& 1.62E-01 & 9.04E-02 & 8.87E-02 \\
 $\Delta x/\eta$ & 1.51 & 1.22 & 1.55 \\
 $\delta_L/\Delta x$ & 19.95 & 136.2 & 35.3 \\
 $\delta_t$ [$\mu$s] & 2E-3 & 1E-03 & 2E-03 \\
 \hline
 \end{tabular}
 \caption{\label{Table_1}Details of the parameters for the three 3D DNS cases investigated in this paper. For all the cases, $P$ = 1 atm, $\delta_L=(T_b^\circ - T_u)/|\grad T|_{max}$.$^*$The approximate values of $\lambda_c$ are estimated from \cite{berger2022,lapenna2021}.}
 \end{center}
 \end{table}
 
The parameters of the DNS cases comprising of statistically planar, lean H$_2$-air turbulent premixed flame at atmospheric pressure are presented in Table~\ref{Table_1}. Le05Ka100 and Le1Ka100 represent new datasets computed using an open-source reacting flow DNS solver called the Pencil Code \cite{Collaboration_2021, chaudhuri2015life, dave2018, dave2020}. Case Le08Ka1000 was previously generated and investigated \cite{song2020dns,yuvraj2022local, yuvraj2023} (named as F2 in these references) but included here for additional insights. Le08Ka1000 was simulated using a detailed chemical mechanism from \citet{burke2012comprehensive} comprising of 9 species and 23 reactions. All the standard laminar values are obtained from Chemkin-Premix calculations.  It is to be noted that for the leanest mixture case Le05Ka100, while the standard laminar values obtained are justifiably treated as reference values, they cannot be realistically obtained from experiments due to intrinsic thermo-diffusive instability.  
The cases vary in equivalence ratio, $\phi$ and consequently the Lewis number~\cite{matalon_2003}, $Le$ which is given by:
\begin{equation} \label{eq:Le_eff}
  Le=\begin{cases}
     \frac{Le_O +\mathcal{A} Le_F}{1+\mathcal{A}} & \phi <1,\\
     \frac{Le_F +\mathcal{A} Le_O}{1+\mathcal{A}} & \phi>1,
  \end{cases}
\end{equation}
where
\begin{equation}
  \mathcal{A}=\begin{cases}
     1+\beta(\phi^{-1}-1) & \phi <1,\\
     1+\beta(\phi-1) & \phi>1.
  \end{cases}
\end{equation}

The combination of selected integral length scale $L_{11}$~\cite{pope_turbulentflows} and root-mean-square of fluctuating velocity $u_{rms}$ result in a Karlovitz number, $Ka\sim\mathcal{O}(100)$ for Le05Ka100 and Le1Ka100 whereas $Ka\sim\mathcal{O}(1000)$ for Le08Ka1000. Here, $Ka=\tau_f/\tau_{\eta}$ is defined as the ratio of flame time scale, $\tau_f = \delta_L/S_L$, and the Kolmogorov time scale $\tau_{\eta}$ in the unburnt mixture. $\delta_L=(T^\circ _b - T_u)/|\grad T|_{max}$ is the standard laminar flame thickness while $T^\circ _b$ is the adiabatic flame temperature and $S_L$ is the corresponding standard laminar flame speed. Reynolds number $Re=u_{rms} L_{11}/\nu \sim 700-1000$ for all the cases. For the present scope of work, the leanest case Le05Ka100 with $Le=0.48$, is of particular interest. 

The Pencil Code \cite{Collaboration_2021} solves the governing equations of mass, momentum, energy and species conservation using a sixth-order central difference scheme to discretize all spatial terms, except for the convective terms, for which a fifth-order upwind scheme is used. A low-storage, third-order accurate Runge-Kutta RK3-2N scheme is used for time marching. A detailed chemical mechanism with 9 species and 21 reactions by \cite{li2004} was used to model the H$_2$-air chemistry for the cases Le05Ka100 and Le1Ka100. The simulation was carried out in two stages for cases Le05Ka100 and Le1Ka100. First, homogeneous isotropic turbulence is generated in a cube comprising of the reactant mixture. The $Re_t$ and $Ka$ reported in Table~\ref{Table_1} are computed for this cube. Next, this turbulent reactant mixture is superimposed on the mean flow and fed through the inlet of a cuboidal domain to interact with a planar laminar premixed flame imposed in the initial field (at $t=0$) as shown in Fig.~\ref{Fig_1}a. The Navier-Stokes characteristic boundary conditions (NSCBC) were imposed in the inflow and outflow direction of the cuboid, with periodic boundary conditions in the transverse directions. The cutoff wavelength $\lambda_c$ for all three H$_2$-air mixtures is presented in Table~\ref{Table_1}. These are estimated from the 2D laminar simulation results from \cite{berger2022, lapenna2021}. For the Le05Ka100 case, we find $L_y/\lambda_c \approx 4$, where $L_y=L_z$ are the transverse dimensions of the cuboid. However, these estimates are from freely propagating laminar flames without any imposed turbulence. At large $Ka$, as in our case, it is unlikely that large cells will have the time to develop before being severely disrupted by the imposed turbulent flow \cite{lipatnikov2005molecular,chaudhuri2011,chomiak2023simple}. Interaction of instability with imposed flow fluctuations have been recently investigated in \cite{lapenna2024, berger2024effects}.

In addition to the 3D-DNS, two additional simulations are performed for each of the DNS cases under similar conditions, for the model analysis. (i) One-dimensional simulations for cylindrical, laminar, premixed, inwardly propagating flames (IPF) using the Pencil Code. The detailed methodology followed for the IPF simulation has been discussed by ~\citet{yuvraj2023}. (ii) Symmetric, laminar, premixed, counter flow flames (CFF)~\cite{law2006} with reactant mixtures entering from both inlets at varying inlet velocities 
using Chemkin. It is ensured that both solvers produce identical solutions for the standard laminar case.

In the present study, we employ isotherms to represent the flame surface within the flame structure. Thus the local flame-displacement speed $S_d$ is evaluated from the DNS fields using temperature as an iso-scalar surface from the right-hand side of the energy equation (Eq.~(\ref{Eq_2})).  
The temperature-based progress variable $c$ used is defined as $c = (T-T_u)/(T_b^\circ -T_u)$, where $T_b^\circ$ is the adiabatic flame temperature. While the actual burned gas temperature of the lean flames may be different from $T_b^\circ$, considering that the temperature on an iso-mass fraction surface may change even more and that the flame speed depends most strongly on temperature \cite{law2006}, temperature-based $c$ is adopted as in \cite{chaudhuri2017}. It has been ascertained that results with the progress variable based on hydrogen mass fraction ($Y_{H_2}$), $c_Y=1-Y_{H_2}/Y_{H_2,u}$, are qualitatively similar to all the results presented in this paper. Recognizing the caveats presented in \cite{howarth2022empirical} due to super-adiabatic temperatures, the analysis is restricted to $c \leq 0.8$. 

All three DNS are computed with mixture averaged diffusivity neglecting the Soret effect as in \cite{howarth2022empirical, howarth2023thermodiffusively}. Soret effect has been shown to induce non-trivial effects on the local consumption speed in lean hydrogen-air flames, especially at positive curvature \cite{schlup2018validation}. Although the differences are quantitative, the qualitative behavior remains the same \cite{song2022diffusive, aspden2011,grcar2009}.

\section{Results and Discussion}
\label{sec: 3}

\begin{figure}[ht!]
\centering
\includegraphics[trim=0cm 0cm 0.cm 0cm,clip,width=390pt]{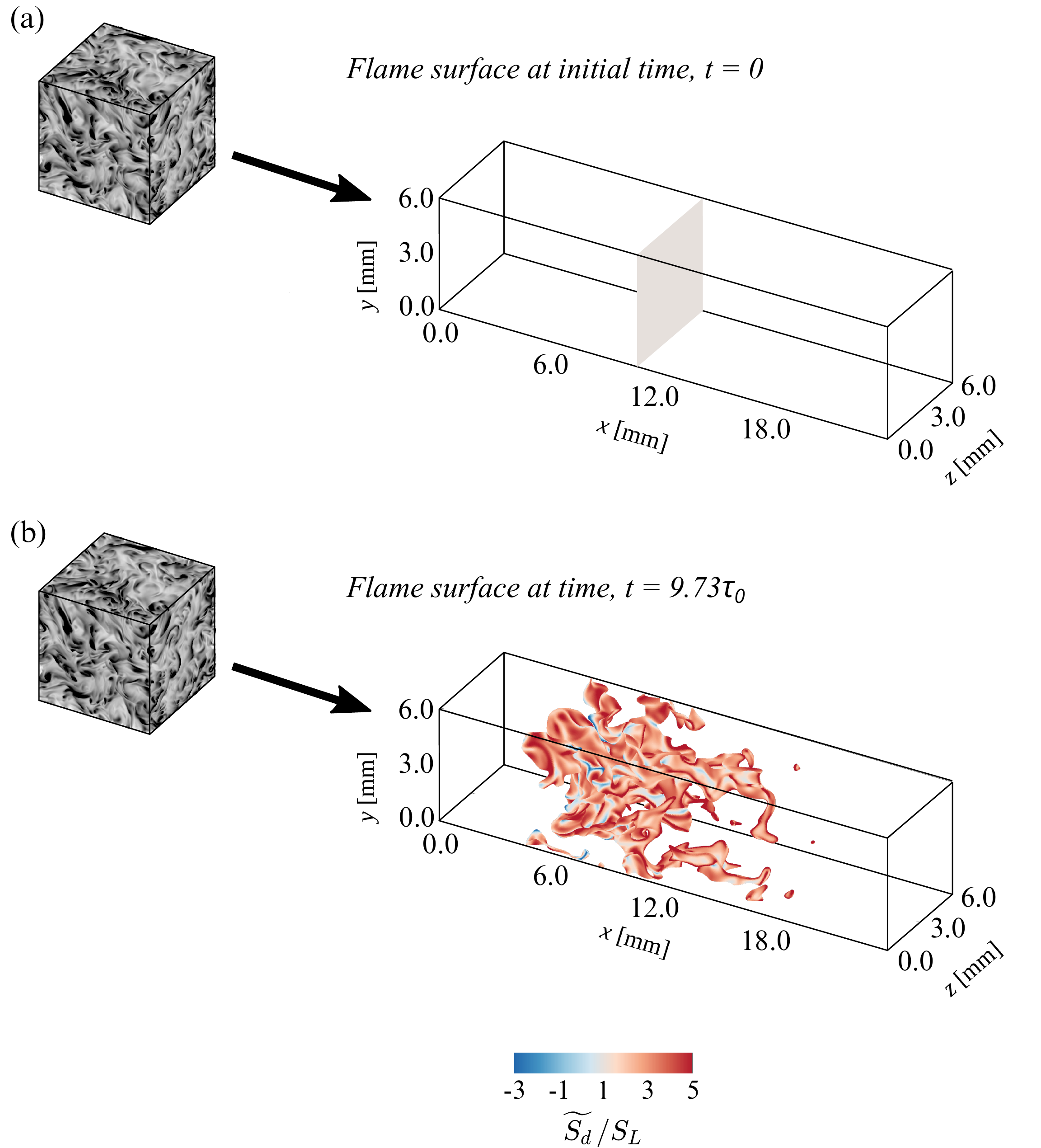}
\vspace{10 pt}
\caption{Flame surface ($c_0=0.2$) for Le05Ka100 at (a) $t=0$ (b) $t=9.73\tau_0$ colored by $\widetilde{S_d}/S_L$, where the integral time scale, $\tau_0=L_{11}/u_{rms}$. The cube containing homogeneous isotropic turbulence fed through the inlet boundary is also shown.}
\label{Fig_1}
\end{figure}

\begin{figure*}[ht!]
\centering
\includegraphics[trim=0cm 0cm 0cm 0cm,clip,width=390pt]{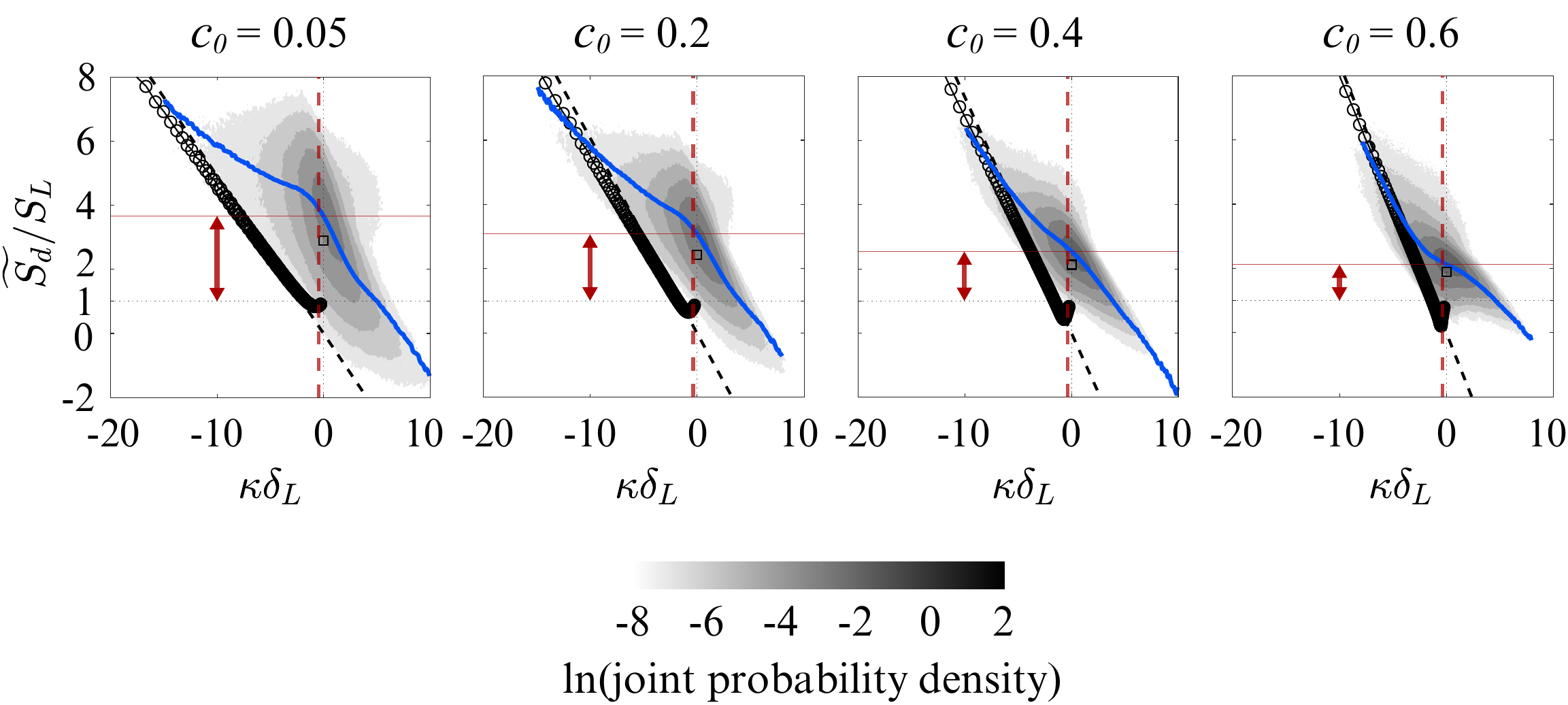}
\vspace{10 pt}
\caption{Joint probability density function (JPDF) of normalized density-weighted flame displacement speed, $\widetilde{S_d}/S_L$, and normalized curvature, $\kappa\delta_L$, for iso-scalar, $c_0=0.05, 0.2, 0.4$ and $0.6$ for Le05Ka100. The colorscale represents the natural logarithm of the JPDF magnitude. The solid blue curve is the conditional mean, $\langle\widetilde{S_d}|_{\kappa}\rangle/S_L$, at a given $\kappa$, obtained from the DNS. The solid horizontal and the dashed vertical red lines show $\widetilde{S_{d,0}}/S_L$ and $\langle\kappa\delta_L\rangle$, respectively. In this paper, we calculate the mean of the variables conditioned on $\kappa=0$ over a narrow range $-0.1<\kappa\delta_L<0.1$. The dashed black line represents the analytical model, $\widetilde{S_d}=-2\widetilde{\alpha_0}\kappa$ and the solid black curve with circular markers represent the IPF results. The hollow square symbol marks $\widetilde{S_{d,0}}/S_L$ conditioned on zero tangential strain rate, $\widetilde{S_{d,0}}|_{a_T=0}/S_L$.}
\label{Fig_2}
\end{figure*}

Figure~\ref{Fig_1}b presents the iso-surface $c=c_0$ at time $t=9.73\tau_0$ for the case Le05Ka100. Here $c_0=0.2$ and $\tau_0=L_{11}/u_{rms}$. The surface is colored with normalized density-weighted flame displacement speed, $\widetilde{S_d}/S_L=\rho S_d / \rho_u S_L $. Qualitatively, an enhancement in $\widetilde{S_d}$ up to a factor of five over $S_L$ is observed at the large negatively curved trailing edge of the flame surface due to flame-flame interaction. Le05Ka100 shows $\widetilde{S_d}/S_L>1$ for most parts of the iso-scalar surface, in contrast to cases with $Le\approx1$ discussed in the literature~\cite{yuvraj2022local, yuvraj2023} for which most of the regions on the flame surface have $\widetilde{S_d}\approx S_L$. 

The widespread enhancement of average $\widetilde{S_d}$ over $S_L$ for Le05Ka100 is also evident from Fig.~\ref{Fig_2}, which presents the joint probability density function (JPDF) of normalized density-weighted local flame displacement speed, $\widetilde{S_d}/S_L$ and the non-dimensional curvature, $\kappa\delta_L$ for Le05Ka100 at $c_0=0.05$, $0.2$, $0.4$ and $0.6$ over multiple time instances. The blue curve shows the conditional mean of the normalized density-weighted local flame displacement speed at a given curvature $\langle\widetilde{S_d}|_{\kappa}\rangle/S_L$. Henceforth, $\langle \boldsymbol{\cdot} \rangle$ denotes the mean of a quantity over all the points corresponding to an iso-surface defined by $c_0$. The vertical dotted red lines denote the mean non-dimensional curvature $\langle\kappa\delta_L\rangle$ and indicate that $\langle\kappa\delta_L\rangle \approx 0$ for all $c_0$. Figure~\ref{Fig_2} overlays the IPF simulation results in solid black lines with black circular symbols, and the corresponding analytical model Eq.~(\ref{Eq_Sd_I}) in dashed black line. $\widetilde{\alpha_0}$ is the density-weighted thermal diffusivity computed using Chemkin-Premix~\cite{kee1985premix}. The horizontal dotted black line denotes the normalized standard laminar value $\widetilde{S_d}/S_L=1$. The horizontal solid red line indicate the magnitude of $\widetilde{S_{d,0}}/S_L=\langle\widetilde{S_d}|_{\kappa=0}\rangle/S_L$.
It should be noted that the mean of the variables (including $\widetilde{S_d}$) conditioned on zero-curvature is calculated over a narrow range $-0.1<\kappa\delta_L<0.1$ throughout the present study. The individual JPDFs also include a hollow square marker in black, which shows the mean of $\widetilde{S_d}|_{\kappa=0}/S_L$ conditioned on zero mean tangential strain rate. This is discussed later in the paper. 

The two notably overarching observations from Fig.~\ref{Fig_2} are the following:

(i) The conditional mean of the JPDF (blue curve) asymptotically matches with the IPF simulation as well as with the analytical model prediction (Eq. \ref{Eq_Sd_I}) for $\kappa \delta_L \ll -1$ for all $c_0$. In fact, with the increase in $c_0$, the conditional mean aligns perfectly with the IPF and the model over a larger $\kappa \delta_L$ range. The slope of $\langle \widetilde{S_d}|_{\kappa}\rangle/S_L$ in the asymptotic limit appears to increase with $c_0$, consistent with the IPF simulation and model prediction. The increase in slope is mainly due to the increasing density-weighted thermal diffusivity with $c_0$, $\widetilde{\alpha_0}$ in the model. 

(ii) The net enhancement of mean $\widetilde{S_d}$ at zero curvature, over $S_L$, is quantified by deviation of $\widetilde{S_{d,0}}/S_L$ above unity and is shown by the red arrow. This indicates that the $\widetilde{S_{d,0}}$ is enhanced to around four times $S_L$ for $c_0=0.05$. This deviation of $\widetilde{S_{d,0}}$ from $S_L$ decreases with $c_0$ as we move further into the flame structure. This observation is in contrast with higher $Le \approx 0.7 - 1.0$ cases \cite{dave2020, yuvraj2022local,chaudhuri2023turbulent,yuvraj2023},  where $\widetilde{S_{d,0}} \approx S_L$ for most $c_0$ values \cite{yuvraj2023}, but similar to the observation in \cite{chu2022}. The significance of $\widetilde{S_{d,0}}$ stems from the fact that, on average, the flame surfaces have near zero-curvature (as shown by the dashed vertical red line) and hence in-depth understanding of this deviation of $\widetilde{S_{d,0}}$ from $S_L$ is required. 


\subsection{Behavior of \texorpdfstring{$\widetilde{S_d}$}{} 
during flame-flame interaction at 
\texorpdfstring{$\kappa\delta_L\ll-1$}{}}\label{sec: 3.1}

\begin{figure}[ht!]
\centering
\includegraphics[trim=0cm 0cm 0cm 0.0cm,clip,width=390pt]{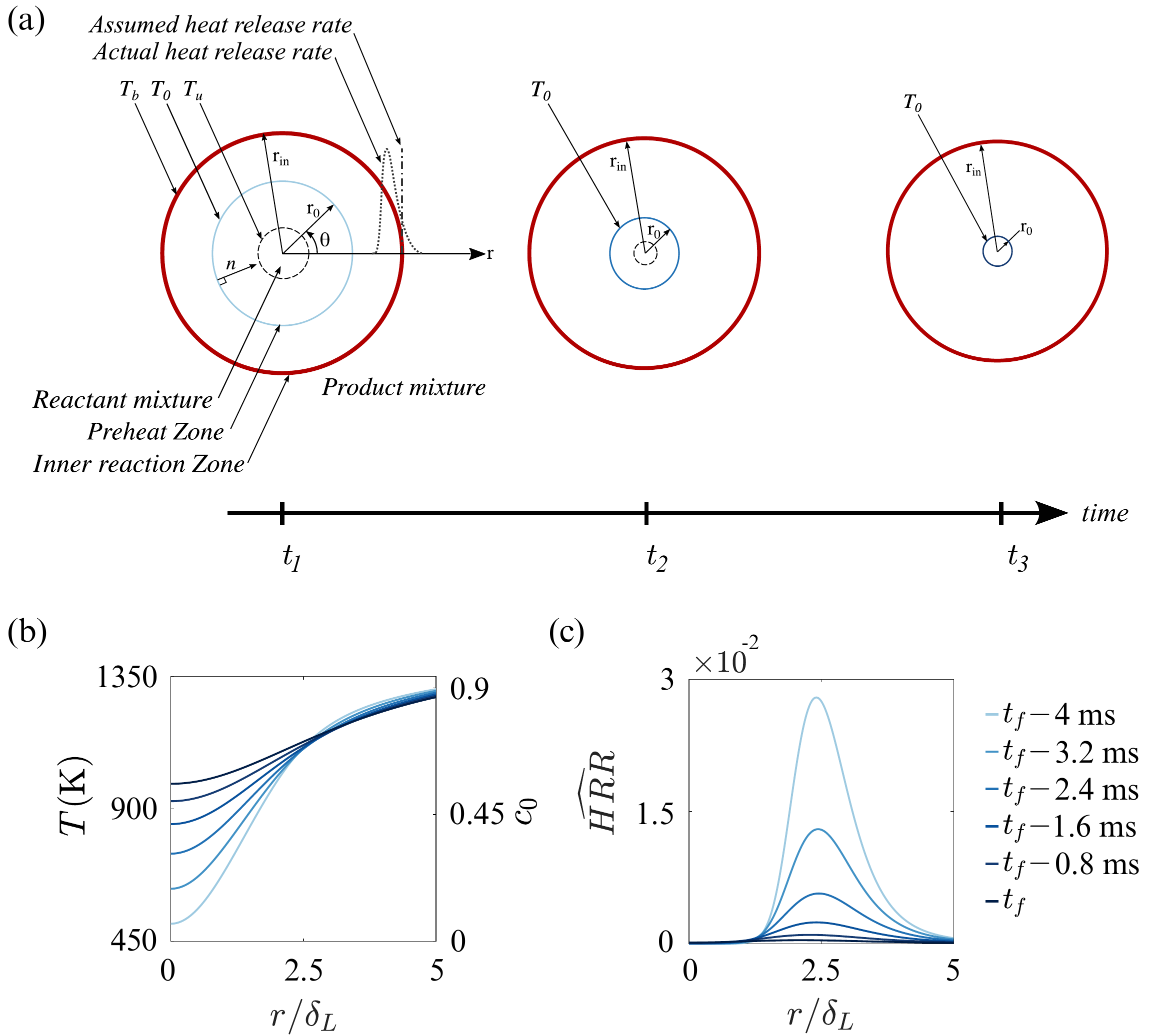}
\vspace{10 pt}
\caption{(a) Schematic of an unsteady IPF showing the isotherm, $T_0$ at different time instances ($t_3 > t_2 > t_1$) during flame-flame interaction. $T_u$ is the unburnt isotherm of radius $r_u$ whereas $T_{in}$ is an isotherm lying in the reaction zone with radius $r_{in}$. (b) Radial temperature (c) normalized heat release rate profiles obtained from IPF for conditions as that of Le05Ka100 at various time instances during flame-flame interaction until the iso-scalar $c_0=0.6$ is annihilated.}
\label{Fig_3}
\end{figure}

The significance of cylindrical, inwardly propagating flame structures in understanding flame-flame interaction at large negative curvatures has been discussed in the Introduction. Figure~\ref{Fig_3}a presents a schematic showing an isotherm $T_0$ within an IPF structure ($T_u < T_0 < T_b^\circ$) at different time instants during flame-flame interaction. This self-interaction of the flame structure is a highly transient phenomenon, as indicated by the fast-decreasing radius of the isotherm of interest ($r_0$). For further details, readers are referred to \citet{yuvraj2023}. To explain the behavior of $\widetilde{S_d}$ at large negative curvature, $\kappa\delta_L \ll-1$ (Fig.~\ref{Fig_2}), the temporal evolution of temperature and the heat release rate (HRR) profiles obtained from the IPF simulation at the thermodynamic conditions of Le05Ka100 during the annihilation event is shown in Fig.~\ref{Fig_3}b and c. $\widehat{HRR}$ is the HRR normalized by its maximum standard laminar value. Note that the HRR barely exists with $3 \%$ of the laminar value even before the onset of interaction due to strong negative stretch rates and $Le$ effects. As the preheat zone flame-flame interaction proceeds, the HRR eventually becomes zero. Thus the heat release zone never interacts. This aligns with the inherent assumption of non-interacting heat release layers at large negative curvatures in the interacting flame theory \cite{dave2020,yuvraj2023}. Consequently, a near-perfect prediction of the DNS results of enhanced $\widetilde{S_d}$ at very large negative curvatures is obtained from the theoretical model (Eq.~\ref{Eq_Sd_I}) for the ultra-lean hydrogen-air flame.

\subsection{Can \texorpdfstring{$\widetilde{S_d}$}{} 
at 
\texorpdfstring{$\kappa\delta_L=0$}{} be described using counterflow flames?}\label{sec: 3.2}

Previous sub-sections showed that a 1D cylindrical flame model could successfully explain the limiting behavior of $\widetilde{S_d}$ at large negative $\kappa$ for locally interacting premixed flames in turbulence. It is well known that the material or propagating surfaces undergo straining in turbulence \cite{yeung1990,pope1989curvature,pope1988,girimaji1992}. Hence, we use a planar flame with finite tangential strain rate to understand the large increase in $\widetilde{S_{d,0}}/S_L$ as found from DNS at $\kappa\delta_L\approx 0$. We compare the DNS results, averaged over zero-curvature regions, with those obtained from the 1D premixed counterflow flame (CFF) \cite{Coulon2023, detomaso2023, hawkes2006b, lee2022_2} at equal average tangential strain rates, i.e., $\langle a_{T}|_{\kappa=0} \rangle = a_{T, CFF}$. For consistency in nomenclature, we define $a_{T,0} = \langle a_{T}|_{\kappa=0} \rangle$ for DNS results, whereas the subscript CFF is used for counterflow flame. The comparative analysis is performed for all cases at the same values of $c_0$.

\begin{figure}[ht!]
\centering\includegraphics[trim=0cm 0cm 0cm 0cm,clip,width=390pt]{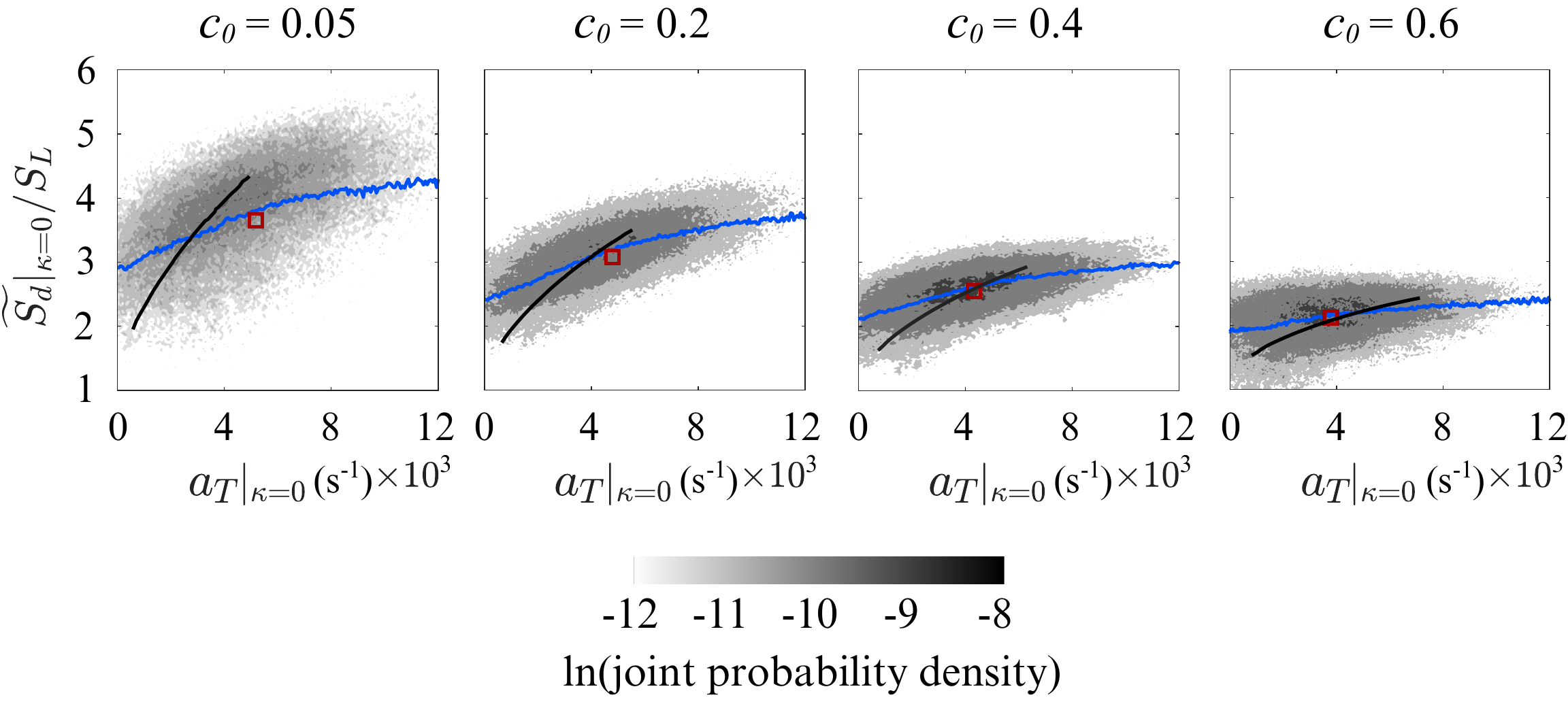}
\vspace{10 pt}
\caption{Joint probability density function (JPDF) of $\widetilde{S_d}|_{\kappa=0}/S_L$ and $a_T|_{\kappa=0}$ for Le05Ka100 at $c_0=0.05, 0.2, 0.4$ and $0.6$. The colorscale corresponds to the natural logarithm of the joint probability density magnitude. The conditional mean of normalized density-weighted local flame displacement speed at zero-curvature given tangential strain, $\langle \widetilde{S_d}|_{\kappa=0, a_T}\rangle/S_L$ is presented by the blue curve. The hollow red square represents the point ($a_{T,0}$, $\widetilde{S_{d,0}}/S_L$). The solid black curves represent the corresponding CFF solution.}
\label{Fig_4}
\end{figure}

\begin{figure}[ht!]
\centering\includegraphics[trim=0cm 0cm 0cm 0cm,clip,width=390pt]{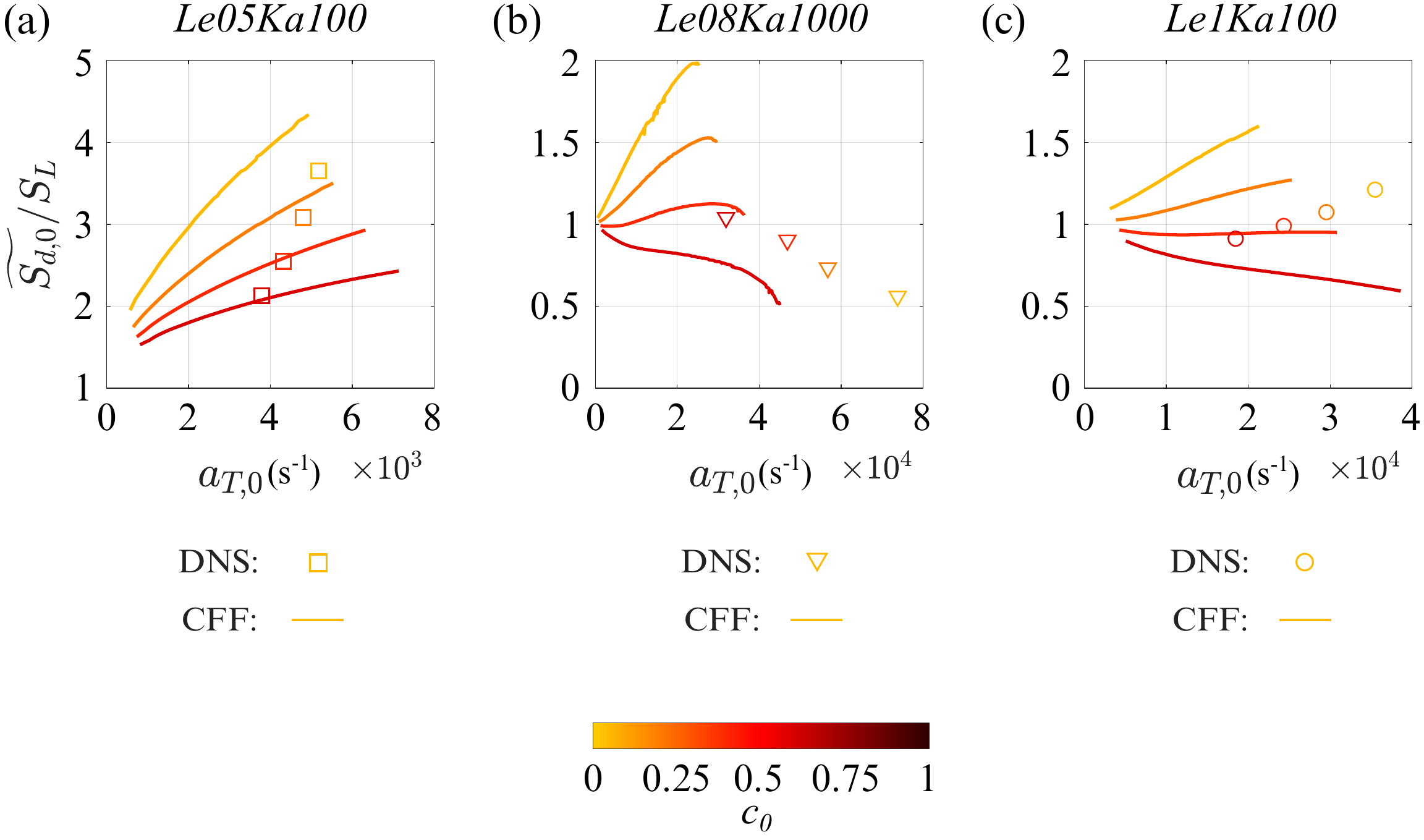}
\vspace{10 pt}
\caption{The variation of $\widetilde{S_{d,0}}/S_L$ with $a_{T,0}$ for (a) Le05Ka100 (b) Le08Ka1000 (c) Le1Ka100 in hollow markers. The solid lines represent the results from CFF. The colorscale corresponds to $c_0$.}
\label{Fig_5}
\end{figure}

Figure~\ref{Fig_4} presents the JPDF of $\widetilde{S_{d}}/S_L$ and $a_T$ both conditioned to $\kappa=0$ for Le05Ka100 at $c_0=0.05, 0.2, 0.4$ and $0.6$. Each of the JPDF is superimposed with $\langle \widetilde{S_d}|_{\kappa=0,a_T} \rangle/S_L$, i.e., the conditional mean of $\widetilde{S_d}/S_L$ given $a_T$, at zero-curvature shown in blue curve. The hollow square marker in red represents the averages ($a_{T,0}$, $\widetilde{S_{d,0}}/S_L$). The CFF solutions are shown in black curves. Apparently, the nature of the overall response of $\langle \widetilde{S_d}|_{\kappa=0, a_T} \rangle/S_L$ to tangential strain rate $a_T|_{\kappa=0}$ is qualitatively similar to that given by CFF \cite{vance2021physical}. However, the strain response of the flame in turbulence is much weaker than that in the CFF configuration. Interestingly, while the CFF solution approaches $S_L$ at zero tangential strain rate, $\langle \widetilde{S_d}|_{\kappa=0, a_T = 0} \rangle > S_L$ for all isotherms. We can recall this observation from (the square markers) Fig.~\ref{Fig_2} as well. Comparing the point ($a_{T,0}$, $\widetilde{S_{d,0}}/S_L$), with the CFF solution it is seen that CFF curve does not extend till $a_{T, CFF}=a_{T,0}$ for $c_0=0.05$ because it fails to provide any solution. It should be noted that the upper limit of CFF solutions is limited by the extinction strain rate at that $c_0$. For $c_0=0.2$, the black curve lies above the square marker suggesting that the CFF response is stronger than that of the turbulent flame at iso-tangential strain rate condition. At $c_0=0.4, 0.6$ the CFF results lie fairly close to that of $\widetilde{S_{d,0}}/S_L$. The variation of $\widetilde{S_{d,0}}/S_L$ with 
$a_{T,0}$ for Le05Ka100 are shown in Fig.~\ref{Fig_5}a in hollow markers. The CFF solutions at $c_0=0.05, 0.2, 0.4$ and $0.6$ from Fig.~\ref{Fig_4} are also included. Similar plots for Le08Ka1000 and Le1Ka100 are presented in Fig.~\ref{Fig_5}b and c, respectively. The colorscale represents the $c_0$ values. 

Figures~\ref{Fig_5}a and c demonstrate that although the trends of $\widetilde{S_{d,0}}$ with $a_{T,0}$ from DNS are qualitatively similar to that of the CFF solutions they are quantitatively different for both Le05Ka100 and Le1Ka100. Figure~\ref{Fig_5}b reveals a stark contrast between DNS and CFF results for Le08Ka100 across all four values of $c_0$ considered.
It is inferred that under the same tangential strain rate, CFF is unable to describe the average flame response at $\kappa=0$ in 3D turbulent flames. Thus, the canonical configuration falls short in describing the behavior of $\widetilde{S_{d,0}}$. This is consistent with \cite{lee2022_2}, where the local structure of leading regions was compared with those from critically strained flames. 
This was attributed to history or transient effects for turbulent flames by \citet{im1996response}. As such conditioning on $a_T=0$ also yields $\widetilde{S_{d,0}}|_{a_T=0}>S_L$ for Le05Ka100 case as shown using black squares in Fig. \ref{Fig_2} in agreement with \cite{chu2022}.  
Hence, further investigation of $\widetilde{S_{d,0}}$ is necessary based on the mean local flame structure, the effect of which on $S_d$ will be instantaneous, eliminating history effects.  

\begin{figure*}[ht!]
\centering
\includegraphics[trim=0cm 0cm 0cm 0cm,clip,width=390pt]{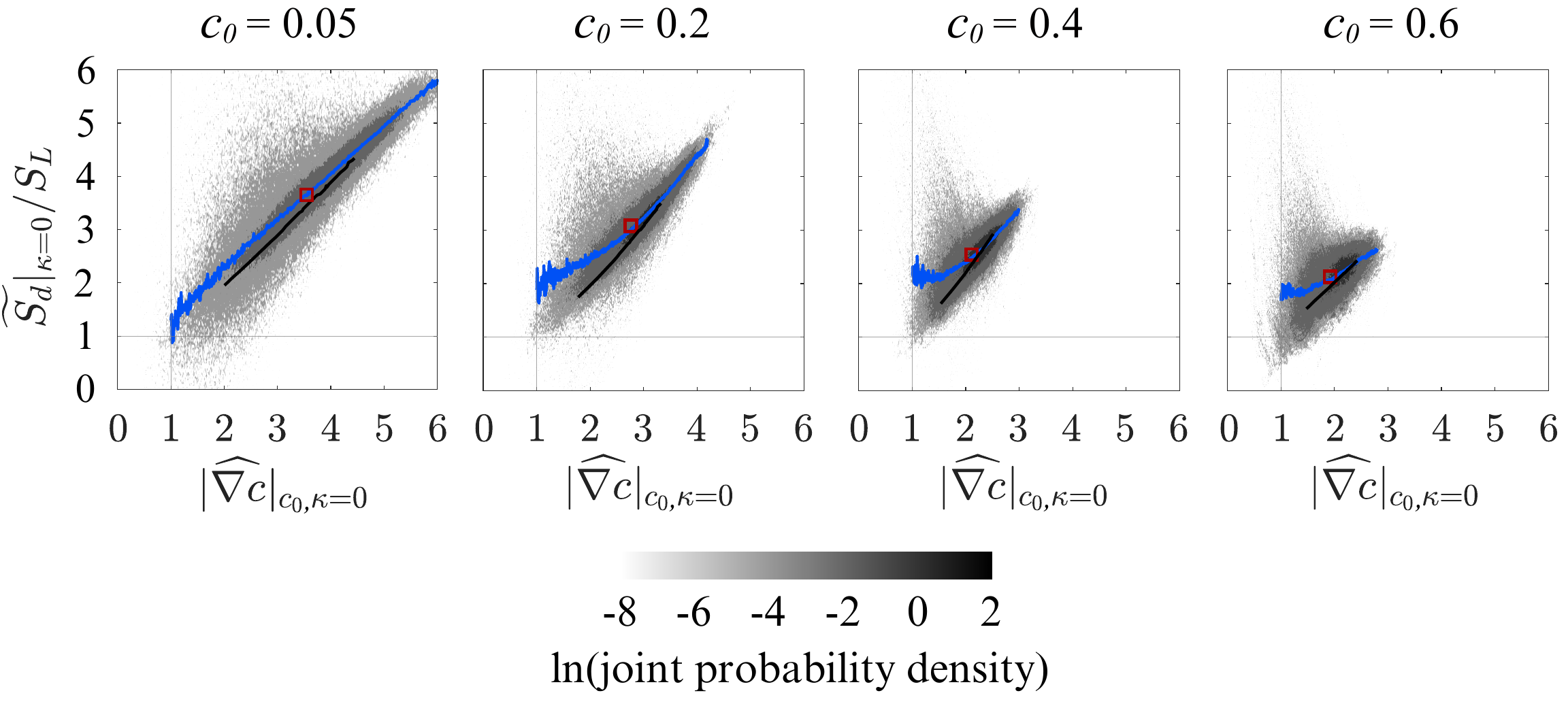}
\vspace{10 pt}
\caption{Joint probability density function (JPDF) of $\widetilde{S_d}|_{\kappa=0}/S_L$ with $|\widehat{\grad c}|_{c_0,\kappa=0}$ for Le05Ka100 at $c_0=0.05, 0.2, 0.4$ and $0.6$. The colorscale corresponds to the natural logarithm of the joint probability density magnitude. The conditional mean given normalized scalar gradient conditioned on zero-curvature, $\langle \widetilde{S_d}|_{\kappa=0,|\widehat{\grad c}|_{c_0}} \rangle$ is presented by the blue curve. The hollow square red markers represent the point ($\langle |\widehat{\grad c}|_{c_0,\kappa=0} \rangle$, $\widetilde{S_{d,0}}/S_L$). The solid black curves represent the CFF solution.}
\label{Fig_6}
\end{figure*}

\begin{figure*}[ht!]
\centering
\includegraphics[trim=0cm 0cm 0cm 0cm,clip,width=390pt]{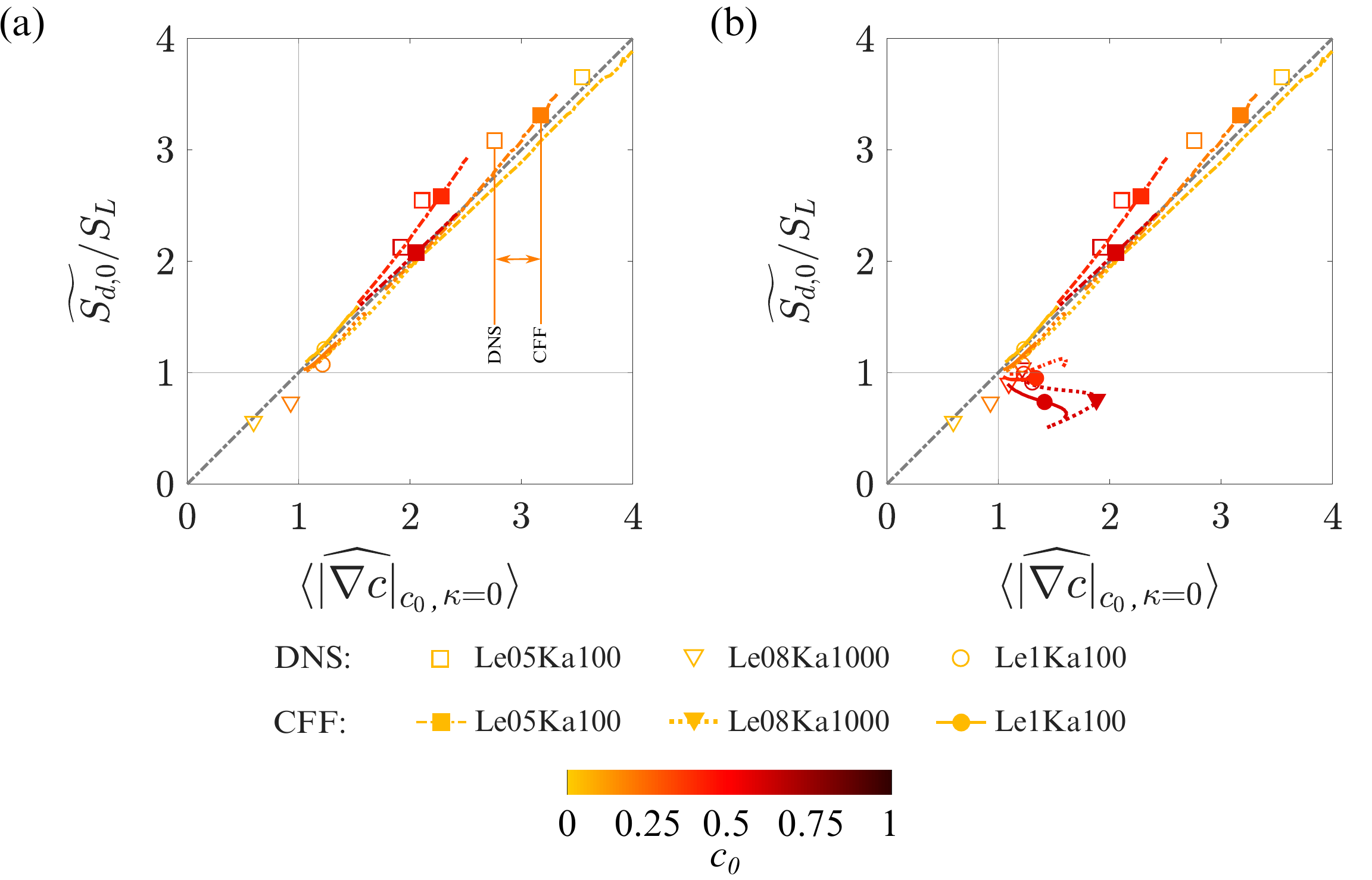}
\vspace{10 pt}
\caption{Correlation of $\widetilde{S_{d,0}}/S_L$ with $\langle|\widehat{\grad c}|_{c_0,\kappa=0}\rangle$ for DNS and laminar CFF (a) in the preheat zone ($c_0$ before maximum heat release rate only) (b) over the entire flame structure at $c_0 = 0.05, 0.2, 0.4$ and $0.6$. Hollow symbols represent DNS data, and filled symbols of the same shape represent CFF data at the same thermodynamic conditions, with  $a_{T, CFF}=a_{T,0}$. Lines show CFF solutions over varying $a_{T,CFF}$.}
\label{Fig_7}
\end{figure*}

\begin{figure*}[ht!]
\centering
\includegraphics[trim=0cm 0cm 0cm 0cm,clip,width=390pt]{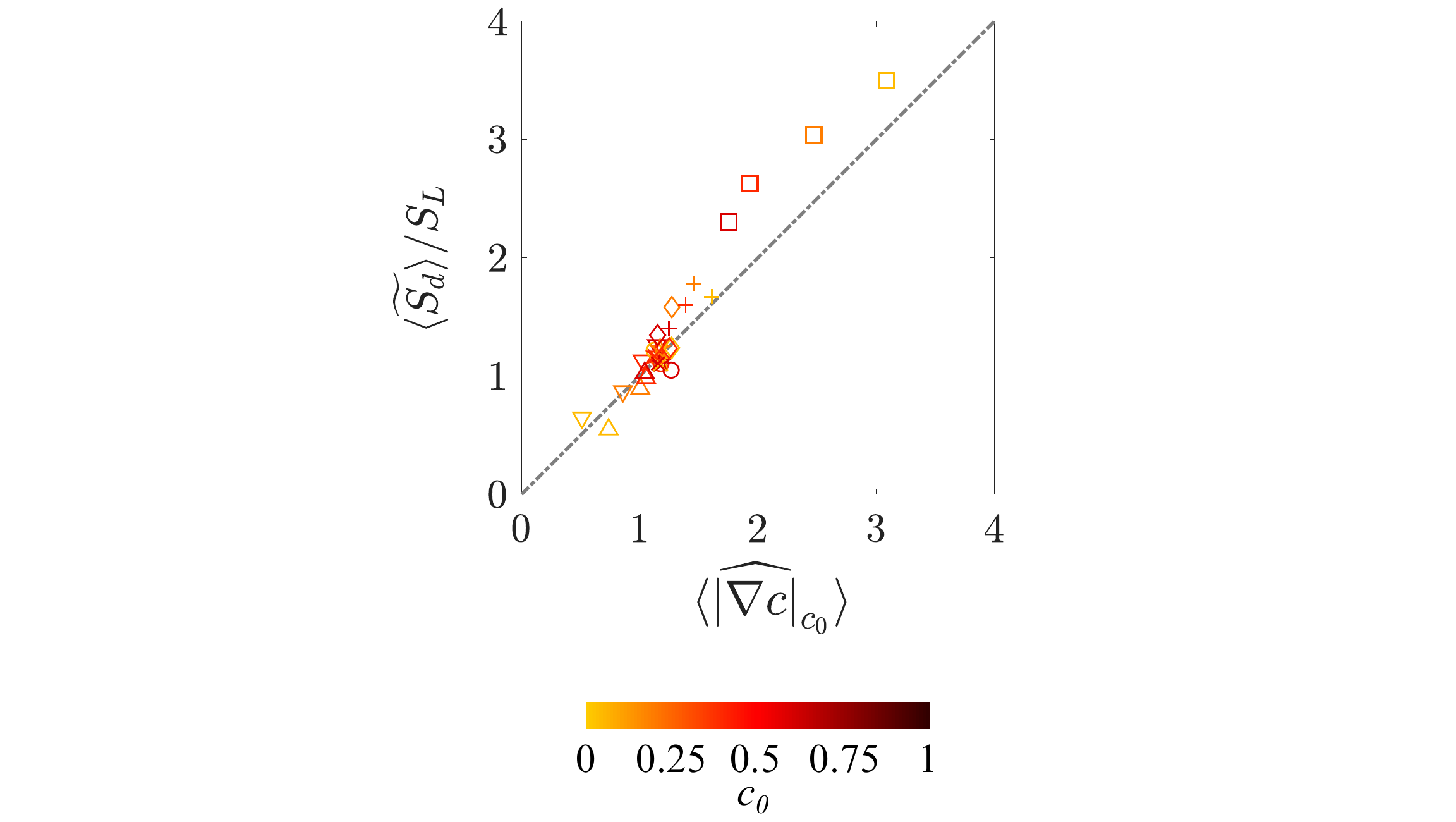}
\vspace{10 pt}
\caption{Correlation of $\langle\widetilde{S_d}\rangle/S_L$ with $\langle|\widehat{\grad c}|_{c_0}|\rangle$ for the present cases and those from previous studies \cite{song2020dns, yuvraj2022local, yuvraj2023}. Data from these additional cases are denoted by: F1: `$\medstar$', F3: `$\times$', F4: `$\medtriangleup$', P3: `$\medlozenge$', P7: `$+$'.}
\label{Fig_8}
\end{figure*}

Previously, the magnitude of the scalar gradient $|\grad c|$ has been used as a measure of localized separation between the iso-scalar surfaces \cite{kim2007, hamlington2011, chaudhuri2017}. The inverse of the local scalar gradient magnitude has been defined as the local flame width \cite{kim2007}. Hence, it is first recognized that the magnitude of the scalar gradient of the progress variable normalized by the corresponding laminar value at that $c_0$, $|\widehat{\grad c}|_{c_0}=|\grad c|_{c_0}/|\grad c|_{c_0, L}$, is a possible measure of the local flame structure relative to its standard laminar counterpart. Thus we seek to correlate $\widetilde{S_d}|_{\kappa=0}/S_L$ and the normalized absolute gradient of the progress variable conditioned on $\kappa=0$. Figure \ref{Fig_6} shows the JPDF of $\widetilde{S_d}|_{\kappa=0}/S_L$ and $|\widehat{\grad c}|_{c_0,\kappa=0}$ for the case Le05Ka100 at the four $c_0$ values. The JPDFs are superimposed with the conditional mean, $\widetilde{S_d}|_{\kappa=0,|\widehat{\grad c}|_{c_0}}/S_L$ (shown in blue) and the corresponding CFF solutions (shown in black). It is apparent that the variables are well correlated for both DNS and  CFF. While the conditional mean from DNS closely follows the CFF solution at $c_0=0.05$ and $0.2$ the average slope of the conditional mean deviates from unity at higher $c_0$. Nevertheless, the point ($\langle|\widehat{\grad c}|_{c_0,\kappa=0}\rangle$, $\widetilde{S_{d,0}}/S_L$) shown in hollow red square lie close to the CFF solution for all $c_0$. Next, we proceed to correlate the averages: $\widetilde{S_{d,0}}/S_L$ and $\langle|\widehat{\grad c}|_{c_0,\kappa=0}\rangle$ from DNS as well as from CFF in Fig.~\ref{Fig_7}a for $c_0$ before the maximum heat release rate and Fig.~\ref{Fig_7}b for all the four $c_0$.  The results from CFF simulations with varying tangential strain rates for $c_0=0.05,0.2,0.4$ and $0.6$ are shown in different line types, whereas the solid markers correspond to CFF solutions at iso-tangential strain rate condition at the corresponding $c_0$. The dash-dotted grey line denotes a linear correlation given by Eq.~(\ref{Eq_5}). 

\begin{equation}
\begin{aligned}
\frac{\widetilde{S_{d}}|_{\kappa=0}}{S_{L_{c,0}}} \approx \frac{{|\grad c|_{c_0,\kappa=0}}}{{|\grad c|_{c_0,L}}} = |\widehat{\grad c}|_{c_0,\kappa=0}
\label{Eq_5}
\end{aligned}
\end{equation}

Overall, $\widetilde{S_{d,0}}/S_L$ show a good correlation with $\langle|\widehat{\grad c}|_{c_0,\kappa=0}\rangle$ for all the DNS cases and CFF solutions in the preheat zone. This observation is consistent with that in \cite{yuvraj2023}, but generalizes the findings to a much wider range of $Le$ conditions over large $\langle|\widehat{\grad c}|_{c_0,\kappa=0}\rangle$ values. Figure~\ref{Fig_8} presents the correlation of $\langle\widetilde{S_d}\rangle/S_L$ and $\langle|\widehat{\grad c}|_{c_0}\rangle$ at $c_0=0.05, 0.2, 0.4$ and $0.6$. The figure also includes the results for cases F1, F4, P3 and P7 from previous studies \cite{song2020dns, yuvraj2022local,yuvraj2023} for a wider range of $Ka$, $Re_t$ and $Le$ conditions. $\langle\widetilde{S_d}\rangle/S_L$ and $\langle|\widehat{\grad c}|_{c_0}\rangle$ are also very well correlated since for all the surfaces $\langle\kappa\delta_L\rangle\approx0$ as evident from Fig.~\ref{Fig_2}. 

Nevertheless, it is apparent from Fig.~\ref{Fig_7} that Le05Ka100 case exhibits more thinned preheat zones with increased $\langle|\widehat{\grad c}|_{c_0,\kappa=0}\rangle$, resulting in large $\widetilde{S_{d,0}}/S_L$, while Le08Ka1000 and Le1Ka100 cases show local broadening in preheat zone thickness with $\widetilde{S_{d,0}}/S_L<1$. While the suggested correlations show overall good agreement for all $Le$ conditions under study, there is an additional important issue that needs further discussion. The data points from CFF calculations at iso-tangential strain rate (shown in filled symbols) for the Le05Ka100 case, appear at higher values along both axes when compared to the DNS in Fig.~\ref{Fig_7}a (shown in hollow symbols but of the same shape). This implies that at $\kappa=0$, the turbulent flame segment, subjected to the same average tangential strain rate, experiences relatively less net flame thinning than its steady strained laminar counterpart. It seems that there is some degree of attenuation in the effect of local straining of the turbulent flame. In other words, the flame surface is locally thinned or propagates faster on average compared to the standard laminar flame. At the same time it is thickened or propagates slower compared to the corresponding CFF. For the cases, Le08Ka1000 and Le1Ka100, the net broadening of the flame structure is observed. One may argue that this broadening is due to turbulence mitigating the local flame surface's response to the local tangential strain rate, reducing $\widetilde{S_{d,0}}$ relative to the CFF predictions at iso-tangential strain rate condition. However, it cannot be concluded that for a premixed flame with even lower $Le$ and $Ka$, $\widetilde{S_{d,0}}$ cannot be higher than the corresponding CFF solution. This is discussed later in section \ref{sec: 3.3.2}. However, based on the DNS cases currently investigated, we observe a consistent effect residing in turbulence-flame interaction where the turbulent flames are broadened on average compared to equivalent strained laminar flames. 

\begin{figure}[ht!]
\centering
\includegraphics[trim=0cm 0cm 0cm 0cm,clip,width=390pt]{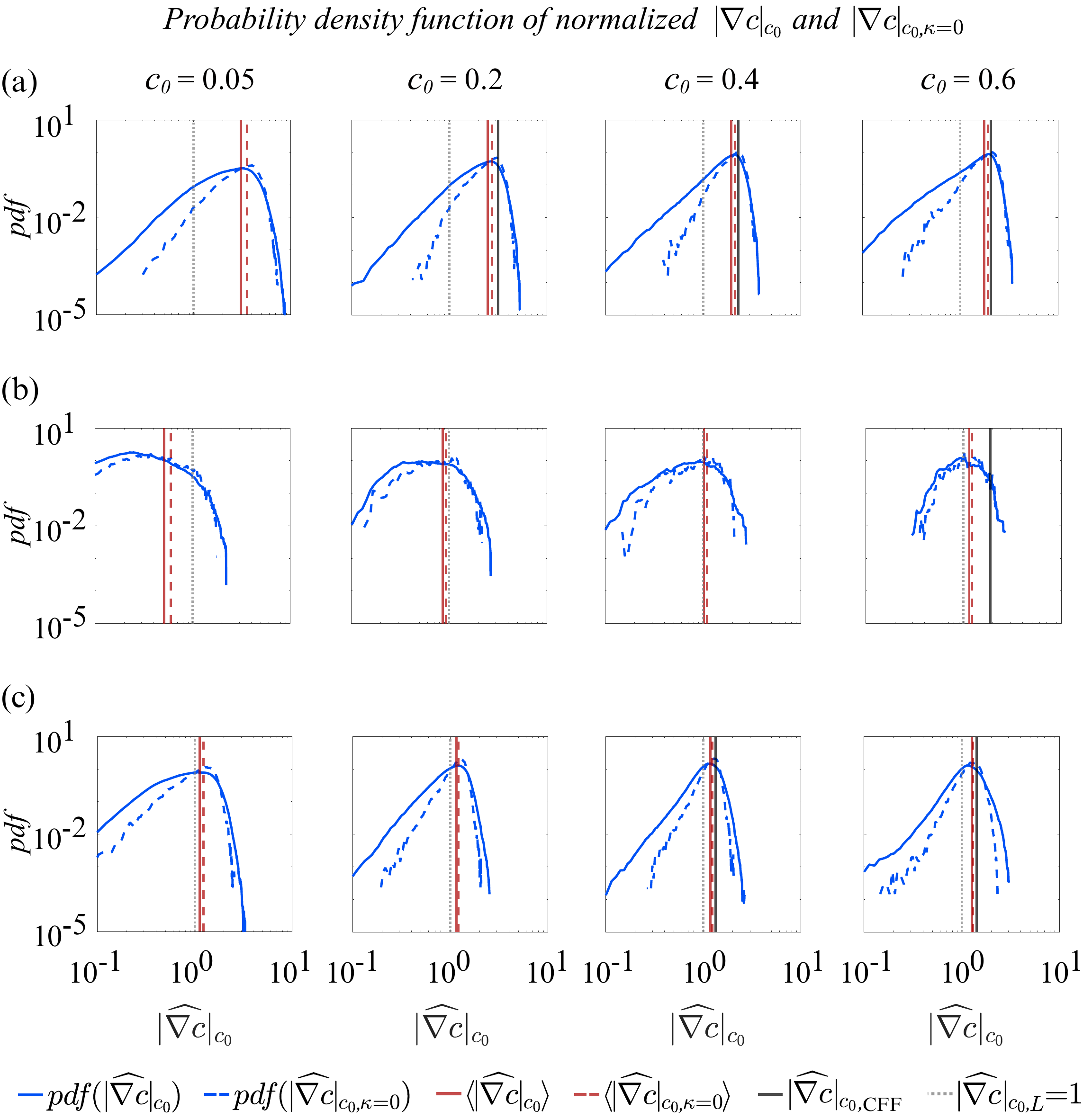}
\vspace{10 pt}
\caption{Probability density function (PDF) of $|\widehat{\grad c}|_{c_0}$ (solid blue curve) and $|\widehat{\grad c}|_{c_0,\kappa=0}$ (dashed blue curve) at $c_0=0.05, 0.2, 0.4$ and $0.6$ for (a) Le05Ka100 (b) Le08Ka1000  (c) Le1Ka100. The solid and dashed vertical red line denote their respective means, $\langle|\widehat{\grad c}|_{c_0}\rangle$ and $\langle|\widehat{\grad c}|_{c_0,\kappa=0}\rangle$. The solid black lines represent the CFF solution, whereas the dotted gray line denotes the standard laminar value, $|\widehat{\grad c}|_{c_0, L}=1$.CFF solutions were not found for some $c_0$; thus, the corresponding PDFs do not contain the vertical black line.}
\label{Fig_9} 
\end{figure}

As a further proof, the probability density function (PDF) of $|\widehat{\grad c}|_{c_0}$ and $|\widehat{\grad c}|_{c_0,\kappa=0}$ is shown in Fig.~\ref{Fig_9} for $c_0=0.05, 0.2, 0.4$ and $0.6$ for the case (a) Le05Ka100 (b) Le08Ka1000 (c) Le1Ka100. The pdfs of $|\widehat{\grad c}|_{c_0}$ and $|\widehat{\grad c}|_{c_0,\kappa=0}$ are shown in solid and dashed blue curves. The solid and dashed vertical red lines represent their corresponding means. The solid black lines are CFF solutions, whereas the dotted gray line denotes the standard laminar value, i.e., $|\widehat{\grad c}|_{c_0,L}=1$. Both the pdfs appear quasi log-normal but with a distinctive negative skewness. The negative skewness in the pdf of $|\widehat{\grad c}|_{c_0}$ was also observed by \cite{hamlington2011, chaudhuri2017}. The negative skewness or occurrence of such small values of $|\widehat{\grad c}|_{c_0}$ is attributed to the regions on the flame surfaces undergoing flame-flame interactions at large negative curvature ($\kappa\delta_L\ll-1$) leading to homogenization of gradients ($|\widehat{\grad c}|_{c_0}\rightarrow0$). On the other hand, pdf of $|\widehat{\grad c}|_{c_0,\kappa=0}$ includes the points on the flame surfaces in the non-interacting regime ($\kappa\delta_L\approx0$) but is still negatively skewed resulting in the corresponding mean less than that of the CFF solution. The existence of such low values of $|\widehat{\grad c}|_{c_0,\kappa=0}$ needs investigation. Overall, on average, the flame for Le05Ka100 and Le1Ka100 for most parts is thinned w.r.t. the standard laminar flame but is thickened w.r.t. the CFF flame structure. However, in Le08Ka1000 the flame is thickened for most parts w.r.t. the standard laminar flame except at large $c_0$ values. 

\subsection{Non-local effects influencing \texorpdfstring{$\widetilde{S_d}$}{} 
at 
\texorpdfstring{$\kappa\delta_L=0$}{}}\label{sec: 3.3}

\subsubsection{Non-local effect of flame-flame interaction at large negative curvatures on \texorpdfstring{$\widetilde{S_d}$}{} 
at 
\texorpdfstring{$\kappa\delta_L=0$}{}}\label{sec: 3.3.1}

\begin{figure}[ht!]
\centering
\includegraphics[trim=0cm 0cm 0cm 0cm,clip,width=390pt]{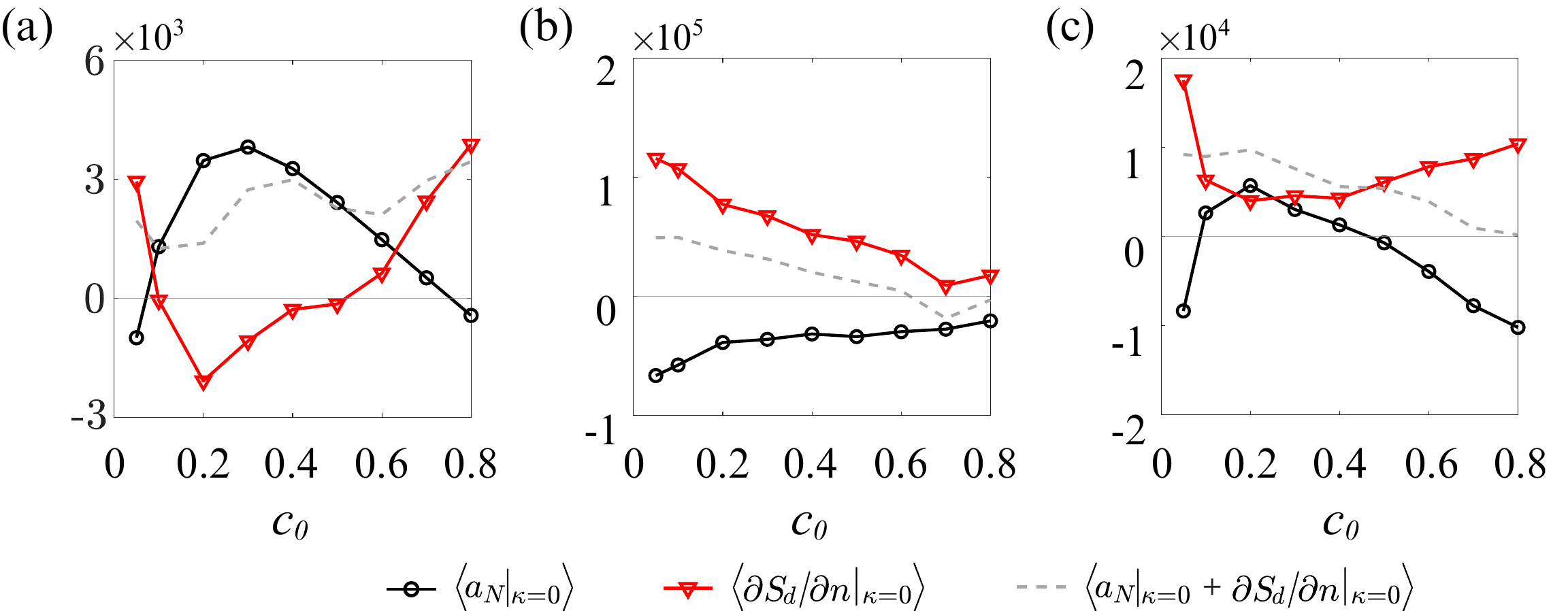}
\vspace{10 pt}
\caption{Variation of mean of normal strain rate ($a_N$) and $\partial S_d /\partial n$ conditioned on $\kappa=0$ with $c_0$ for (a) Le05Ka100 (b) Le08Ka1000 (c) Le1Ka100. The dashed gray curve denotes the sum of the two terms.}
\label{Fig_10} 
\end{figure}

Since $\widetilde{S_{d,0}}$ and $\langle|\widehat{\grad c}|_{c_0,\kappa=0}\rangle$ were found to be well correlated, we seek to understand the deviation of $|\widehat{\grad c}|_{c_0,\kappa=0}$ from its corresponding CFF as well as the standard laminar value. To that end, we consider the transport equation for $|\widehat{\grad c}|$ in the Lagrangian form at the flame surface \cite{sankaran2007,dopazo2015,chaudhuri2017,wang2017_2}:

\begin{equation}
\begin{aligned}
    \frac{\widetilde{D}|\widehat{\grad c}|}{\widetilde{D}t}  = -\left[ a_N + \frac{\partial S_d}{\partial n} \right]|\widehat{\grad c}|
    \label{Eq_9}
\end{aligned}
\end{equation}

\noindent where $a_N$ is the fluid motion-induced normal strain rate and $\partial S_d/\partial n$ is the normal strain rate due to flame propagation. $\boldsymbol{n}=-\grad c/|\grad c|$ is the local normal in the direction of the reactants. Each term on the right-hand side of Eq.~(\ref{Eq_9}) shows the mechanism of thickening or thinning of the flame structure. In steady flames, the sum is identically zero. Figure~\ref{Fig_10} shows the variation of the mean $a _N$ and $\partial S_d/\partial n$ conditioned on $\kappa=0$, with $c_0$ for all the cases. Dopazo et al. \cite{dopazo2015} reported $\langle a_N \rangle>0$ and $\langle \partial S_d/\partial n \rangle <0$ for low $Ka$ flames. In the present study, we found that for Le08Ka1000 and Le1Ka100, $\langle \partial S_d/\partial n |_{\kappa=0}\rangle >0$, similar to those observed in large $Ka$ methane-air Bunsen flames by~\citet{wang2017_2} but in direct contrast to planar laminar unstrained/strained flame behavior where $ \partial S_d/\partial n = -a_N <0  $. Although Le05Ka100 shows $\langle \partial S_d/\partial n |_{\kappa=0}\rangle <0$ for $0.1<c_0<0.5$ its magnitude is much less than $\langle a_N|_{\kappa=0} \rangle$. For $Ka \sim 100$ cases, $\langle a_N|_{\kappa=0} \rangle>0$ possibly due to dilatation, while for $Ka \sim 1000$, $\langle a_N|_{\kappa=0} \rangle<0$. The nature of $\langle a_N|_{\kappa=0} \rangle$ in turbulent premixed flame based on the alignment of the smallest principal strain rate and the normal to the iso-scalar surfaces have been discussed \cite{chakraborty2007influence,chaudhuri2017}. 
Unlike a steady laminar flame, where positive $a_N$ and negative $\partial S_d /\partial n$ balance out, in the present turbulent flames under investigation, their net contribution remains positive (dashed gray curve in Fig.~\ref{Fig_10}). This is primarily due to $\langle \partial S_d/\partial n|_{\kappa=0}\rangle$ attaining positive values or small negative values that are insufficient to balance $\langle a_N|_{\kappa=0}\rangle$, leading to local flame broadening. 

\begin{figure*}[ht!]
\centering
\includegraphics[trim=0cm 0cm 0cm 0cm,clip,width=390pt]{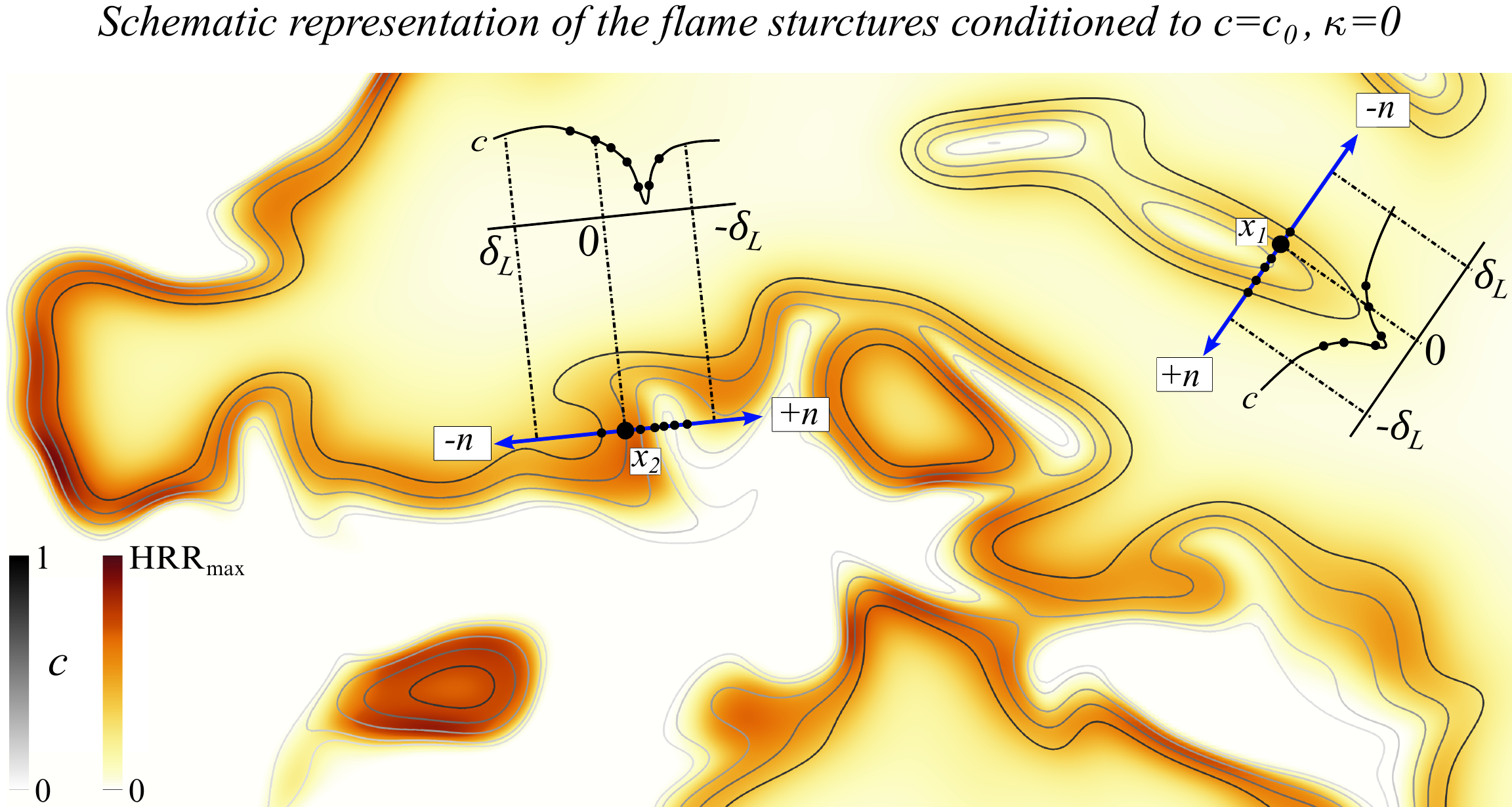}
\vspace{10 pt}
\caption{Two-dimensional contour of heat release rate overlaid with iso-scalar curves at mid-plane extracted from Le05Ka100 at a given time instant. $x_1$ and $x_2$ are the points on a selected iso-scalar surface $c=c_0$ where $\kappa\approx0$. The local normal at these points extending in either direction is shown in blue. The local flame structures based on $c$ extracted on either side of $x_1$ and $x_2$ along their normal direction up to a distance of $\delta_L$ are shown in black. The circular markers on the $c$ profile represent the $c$ values at the corresponding neighboring iso-scalar.}
\label{Fig_11}
\end{figure*}

\begin{figure*}[ht!]
\centering
\includegraphics[trim=0cm 0cm 0cm 0cm,clip,width=390pt]{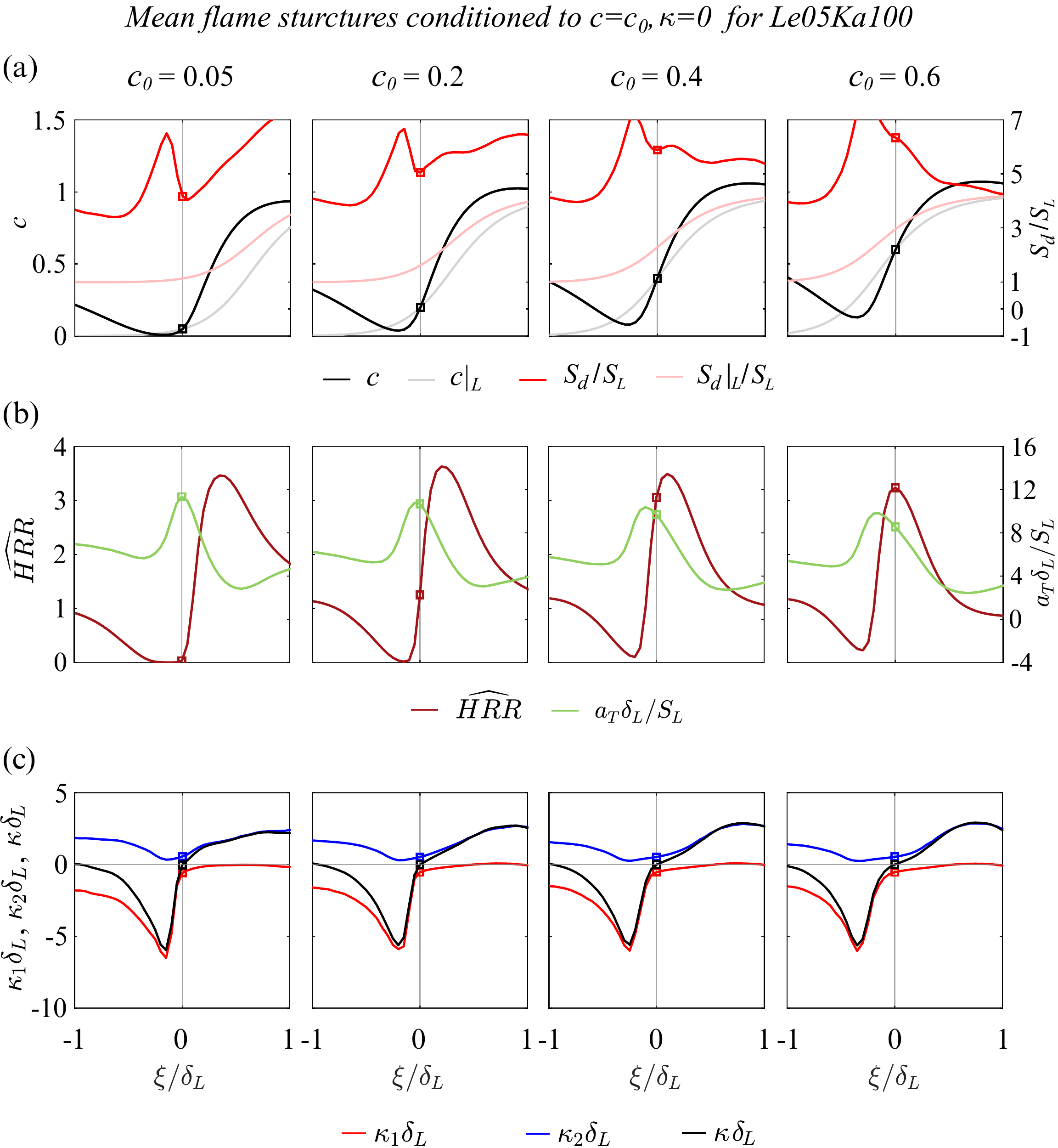}
\vspace{20 pt}
\caption{Mean flame structures of (a) $c$ and $S_d/S_L$ (b) normalized heat release rate, $\widehat{HRR}$ and non-dimensional tangential strain rate, $a_T \delta_L/ S_L$ (c) non-dimensional curvature $\kappa\delta_L$, non-dimensional minimum and maximum principal curvatures, $\kappa_1\delta_L$ and $\kappa_2\delta_L$ conditioned on $\kappa=0$ at $c=c_0$ for Le05Ka100. The faint red and gray curves in the background in (a) represent the corresponding standard laminar flame structures.}
\label{Fig_12}
\end{figure*}

\begin{figure*}[ht!]
\centering
\includegraphics[trim=0cm 0cm 0cm 0cm,clip,width=390pt]{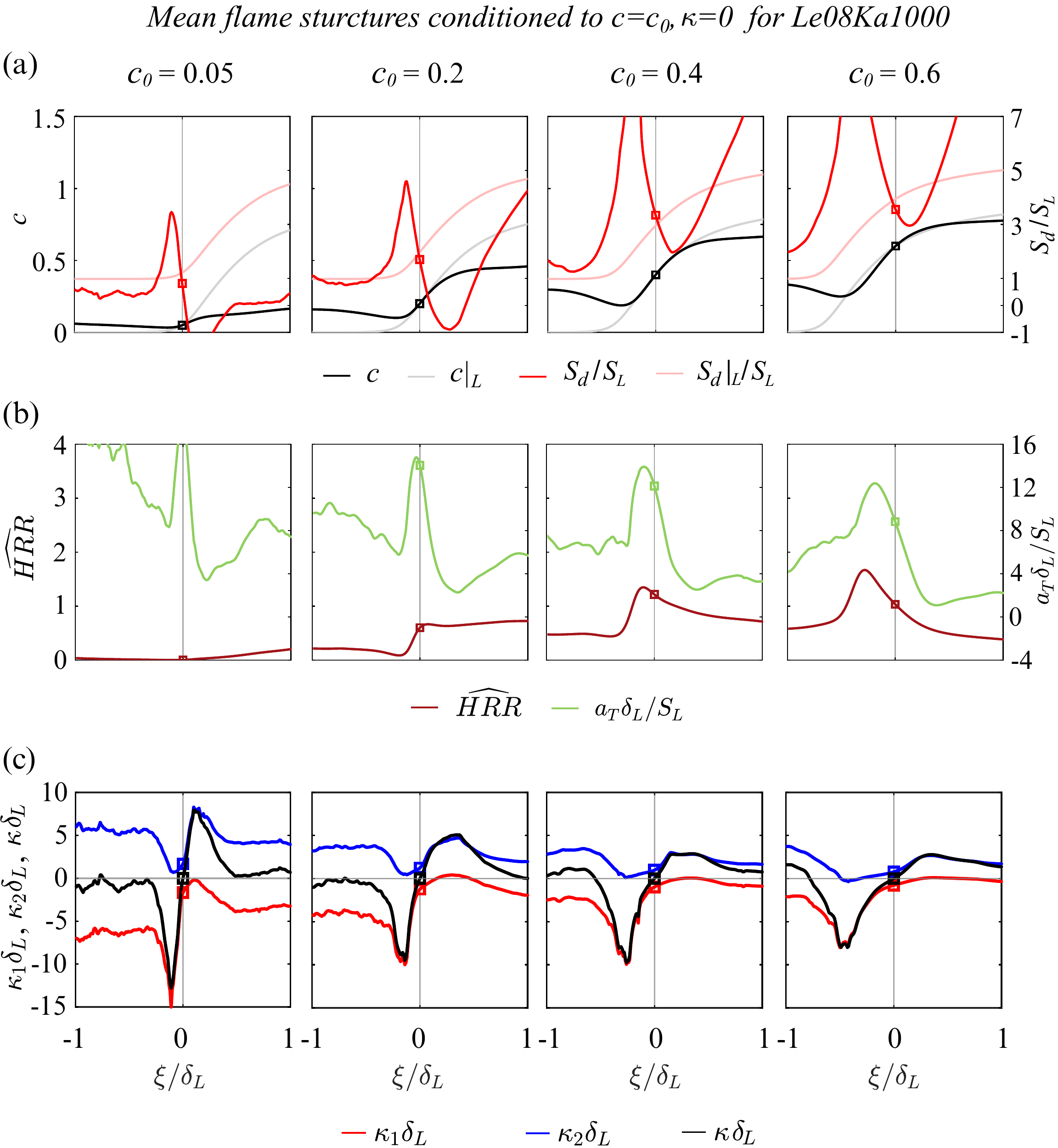}
\vspace{20 pt}
\caption{Mean flame structures of (a) $c$ and $S_d/S_L$ (b) normalized heat release rate, $\widehat{HRR}$ and non-dimensional tangential strain rate, $a_T \delta_L/ S_L$ (c) non-dimensional curvature $\kappa\delta_L$, non-dimensional minimum and maximum principal curvatures, $\kappa_1\delta_L$ and $\kappa_2\delta_L$ conditioned on $\kappa=0$ at $c=c_0$ for Le08Ka1000. The faint red and gray curves in the background in (a) represent the corresponding standard laminar flame structures.}
\label{Fig_13}
\end{figure*}

\begin{figure*}[ht!]
\centering
\includegraphics[trim=0cm 0cm 0cm 0cm,clip,width=390pt]{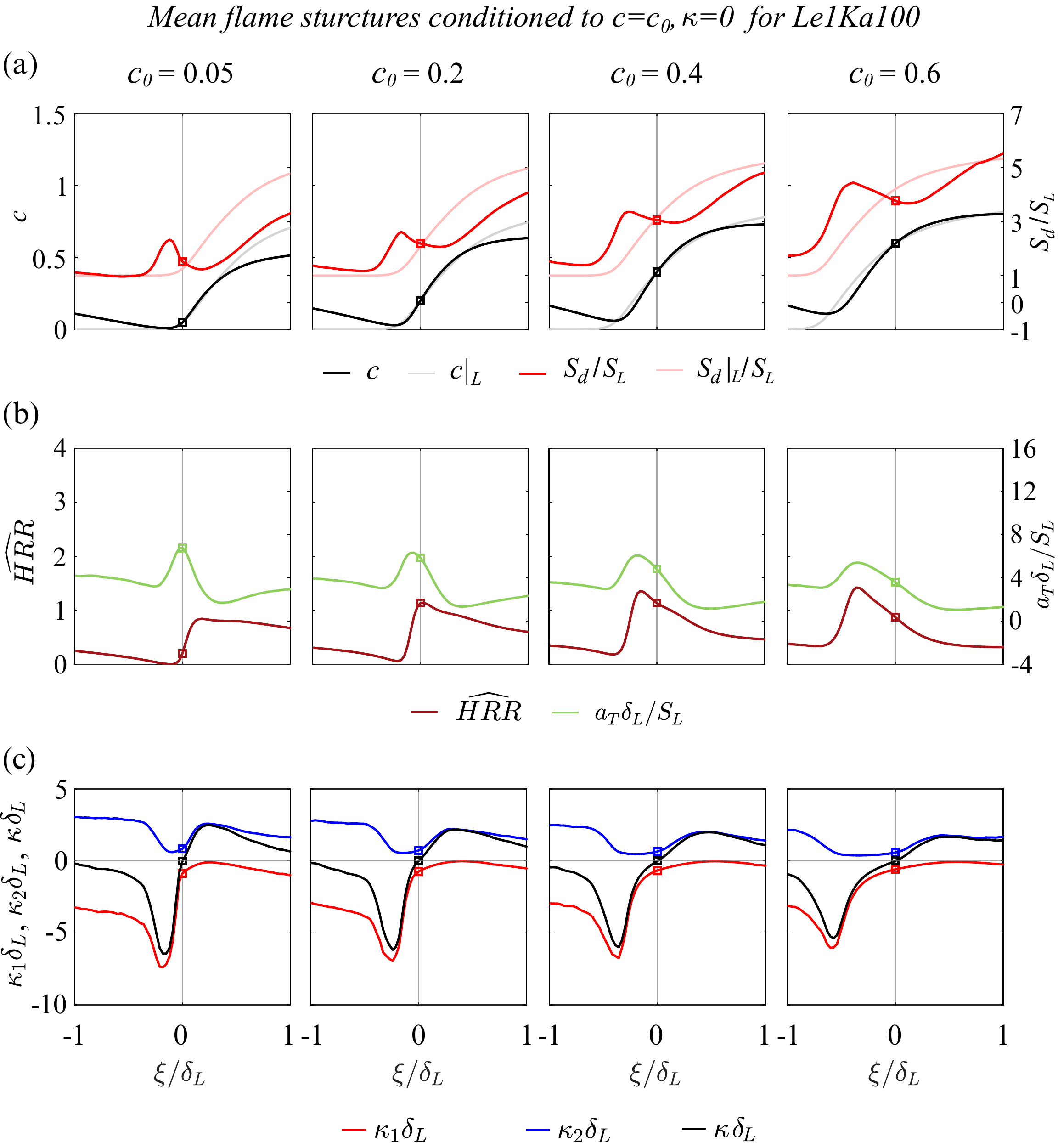}
\vspace{20 pt}
\caption{Mean flame structures of (a) $c$ and $S_d/S_L$ (b) normalized heat release rate, $\widehat{HRR}$ and non-dimensional tangential strain rate, $a_T \delta_L/ S_L$ (c) non-dimensional curvature $\kappa\delta_L$, non-dimensional minimum and maximum principal curvatures, $\kappa_1\delta_L$ and $\kappa_2\delta_L$ conditioned on $\kappa=0$ at $c=c_0$ for Le1Ka100. The faint red and gray curves in the background in (a) represent the corresponding standard laminar flame structures.}
\label{Fig_14}
\end{figure*}

\begin{figure}[ht!]
\centering
\includegraphics[trim=0.cm -1cm 0.cm 0cm,clip,width=390pt]{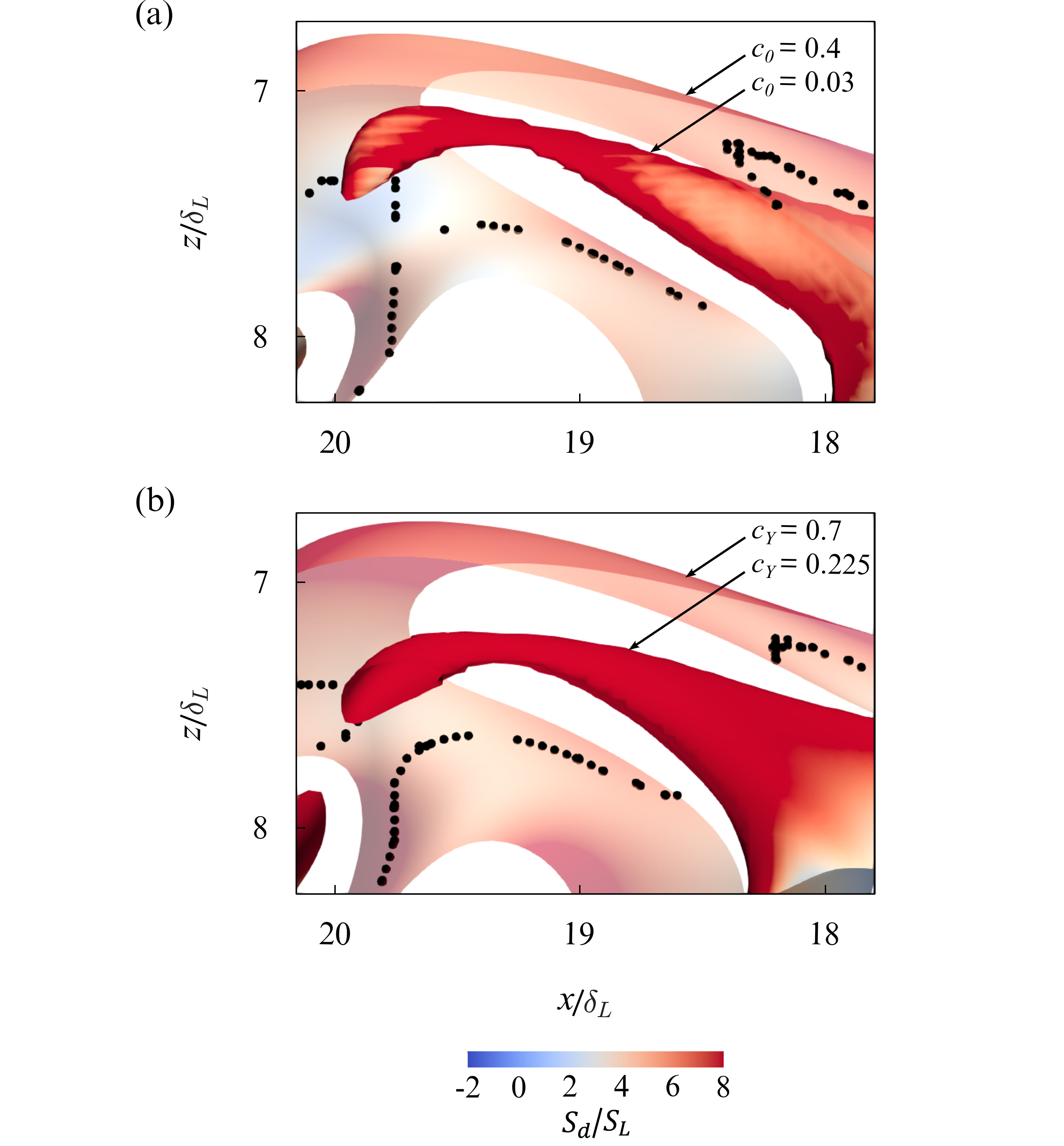}
\vspace{20 pt}
\caption{Segments of the iso-scalar surfaces of Le05Ka100 colored by $S_d/S_L$ based on (a) $T$: $c_0=0.03$ and $0.4$ (b) $Y_{H_2}$: $c_Y=0.225$ and $0.7$. The black markers denote the points with $\kappa=0$ on (a) $c_0=0.4$ and (b) $c_Y=0.7$.}
\label{Fig_15}
\end{figure}

We seek to understand such crucial behavior of the normal gradient of the flame speed, which is central to flame thickening and associated reduction in $\widetilde{S_{d,0}}$. Conditionally averaged flame structures along the local normal directions from surface locations $c=c_0$ where $\kappa=0$, are analyzed. Figure~\ref{Fig_11} presents a schematic including the two-dimensional contour of heat release rate overlaid with iso-scalar curves at mid-plane extracted from Le05Ka100 at a given time instant. The schematic includes two points, $x_1$ and $x_2$ on a selected iso-scalar surface $c=c_0$ where $\kappa\approx0$. The local normal at these points extending in either direction is shown in blue. The flame structures are extracted on either side of $x_1$ and $x_2$ along their normal direction up to a distance of $\delta_L$. The obtained flame structures based on $c$ are shown in black, with the circular markers representing the $c$ values at the corresponding neighboring iso-scalar surface. Note that the actual structures are generated from the 3D DNS fields. For further details regarding the algorithm followed, the readers can refer to~\citet{yuvraj2022local, yuvraj2023}. Figure~\ref{Fig_12}a shows the mean flame structure (black for $c$ and red for $S_d/S_L$) conditioned to $c=c_0$ and $\kappa=0$ for the flame surfaces $c_0=0.05, 0.2,0.4$ and $0.6$ for Le05Ka100. The abscissa $\xi/\delta_L$ denotes the normalized distance along the local normal direction. The corresponding flame structures from standard laminar flame are shown in faint curves. The vertical grey line, i.e., $\xi/\delta_L=0$ is the origin lying on the corresponding $c=c_0$ where $\kappa=0$. The mean $c$ and $S_d/S_L$ values at $\xi/\delta_L=0$ are represented by black and red square markers, respectively. $\partial S_d/\partial n$ is indeed positive or weakly negative at the points on the flame surface conditioned on $\kappa=0$ ($\xi/\delta_L=0$) in the preheat zone in agreement with Fig.~\ref{Fig_10}. This is because such points are preceded by a sharp increase in $S_d$ along $\boldsymbol{n}$ on the upstream, starkly contrasting with any strained or unstrained laminar flame structure. The V-shaped average temperature profile suggesting possible flame-flame interaction just preceding the point of interest is also apparent. The second set of flame structures based on the heat release rate normalized by the corresponding maximum laminar value, $\widehat{HRR}$ (shown in dark red) and non-dimensional tangential strain rate, $a_T \delta_L /S_L$ (shown in green) is included in Fig.~\ref{Fig_12}b. 

Finally, Fig.~\ref{Fig_12}c presents the third set of conditionally averaged flame structures (from $c=c_0$ and $\kappa=0$) along local normal $\boldsymbol{n}$ which includes non-dimensional total curvature, $\kappa\delta_L$ in black, minimum and maximum non-dimensional principal curvatures, $\kappa_1\delta_L$ and $\kappa_2\delta_L$ in red and blue respectively. Coinciding with the peak in $S_d$ and the valley in the $c$ profiles, we observe a sharp drop in average $\kappa$ and $\kappa_1$ profiles just preceding $\xi/\delta_L=0$ with $\kappa_2\approx0$ at large negative $\kappa_1$. This is the quintessential signature of cylindrical flame-flame interaction shown in Fig.~\ref{Fig_2}. This implies that due to their ubiquity at large $Ka$, localized, cylindrical flame-flame interaction precedes the nearly flat locations of the flame surface in the direction of reactants, on average. Since the cylindrical flame-flame interaction enhances $S_d$, the interacting iso-scalar surfaces approach each other faster, in turn separating the trailing surfaces leading to local broadening. The local flame thickening of these nearly flat turbulent flame segments is found to be mainly due to the upstream flame-flame interaction. The results presented in Fig.~\ref{Fig_13}a and c for Le08Ka1000 and Fig.~\ref{Fig_14}a and c Le1Ka100 show qualitatively similar mean structures along the positive normal direction at $\xi/\delta_L =0$. 

Fig.~\ref{Fig_12}b shows that just downstream, for $\xi/\delta_L>0$, the maximum heat release rate of the averaged flame structure is enhanced to around 3.5 times the maximum standard laminar value. This results in an increased temperature gradient at $\kappa=0$ ($\xi/\delta_L=0$), causing the $\widetilde{S_d}$ to increase. The lean hydrogen-air turbulent flames at $\kappa=0$ thus exhibit an enhancement in $\widetilde{S_d}$ at sub-unity $Le$. This will be discussed in detail later in the subsection \ref{sec: 3.3.2}. However, the normalized heat release rate profiles for Le08Ka1000 (Fig.~\ref{Fig_13}b) and Le1Ka100 (Fig.~\ref{Fig_14}b) are different from Le05Ka100 given their $Le$ are close to unity and are also discussed in the subsection \ref{sec: 3.3.2}.  

We visualize the non-local flame-flame interaction upstream of $\kappa=0$ in Fig.~\ref{Fig_15} presenting segments of the iso-scalar surfaces at an instant of time for Le05Ka100 using both isotherms and iso-$Y_{H_2}$ surfaces in (a) and (b), respectively. The iso-scalar surfaces are colored with $S_d/S_L$ as shown by the colorscale at the bottom. The black markers denote the $\kappa=0$ points on $c_0=0.4$ and $c_Y=0.7$ iso-surfaces. From the viewpoint of an observer in these nearly flat yet strained locations (black markers), the flame-flame interaction at the neighboring, highly curved, near cylindrical surfaces increases $S_d$ at those corresponding neighboring surfaces. This causes those neighboring surfaces to separate faster from the reference resulting in flame thickening characterized by diminished scalar gradients.

\subsubsection{Non-local effect of positive curvatures on \texorpdfstring{$\widetilde{S_d}$}{} 
at 
\texorpdfstring{$\kappa\delta_L=0$}{}}\label{sec: 3.3.2}

The previous section discussed how the cylindrical flame-flame interactions upstream (i.e. towards the reactant side) of the zero-curvature regions lead to departure from the local laminar flame structure. Next, we shift our focus to the region downstream (i.e. towards the product side) of the zero-curvature in the direction along the local normal estimated at $\xi/\delta_L=0$. We observe that overall temperature (or $c$) is enhanced for Le05Ka100 (Fig.~\ref{Fig_12}a) resulting in enhanced temperature gradient at $\xi/\delta_L=0$ at all $c_0$. As mentioned before, the peak heat release rate is around 3.5 times the maximum laminar value for Le05Ka100 for all $c_0$ as shown in Fig.~\ref{Fig_12}b. However, for L08Ka1000 with $Le=0.76$, heat release rate normalized by its standard laminar value, $\widehat{HRR}<1$ for $c_0=0.05$ and $0.2$ (Fig.~\ref{Fig_13}b). For $c_0=0.4$ and $c_0=0.6$ the $\widehat{HRR}\approx1.5$. In the preheat zone ($c_0=0.05$ and $0.2$), the diffusion term contributes majorly to $S_d$ while for $c_0=0.4$ and $c_0=0.6$ close to or beyond peak heat release rate the reaction term dominates (see Eq.~(\ref{Eq_2})). In either case, with increasing $c_0$, the difference between the $S_d/S_L$ and $S_d|_L /S_L$ decreases downstream of $\xi/\delta_L=0$. The mean flame structures for Le1Ka100 show similar behavior downstream of $\xi/\delta_L=0$ (Fig.~\ref{Fig_14}a and b). For all the cases, the non-dimensionalized tangential strain rate, $a_T\delta_L S_L$ is positive for $\xi/\delta_L\geq0$. Interestingly, just downstream of $\xi/\delta_L=0$, the flame surfaces are also near cylindrical with maximum principal curvature contributing majorly to the local curvature ($\kappa\delta_L=\kappa_1\delta_L + \kappa_2\delta_L\approx\kappa_1\delta_L, \kappa_2\delta_L\approx0$) for all the cases. It seems that this large curvature and/or the existing tangential strain rate downstream contribute significantly to a large positive stretch. The positive stretching and differential diffusion downstream leads to an enhanced heat release rate, eventually increasing temperature gradient and $S_d$. This contribution from the non-local effect is in addition to the existing tangential strain rate at $\kappa=0$.

\begin{figure*}[h!]
\centering
\includegraphics[trim=0cm 0cm 0cm 0cm,clip,width=390pt]{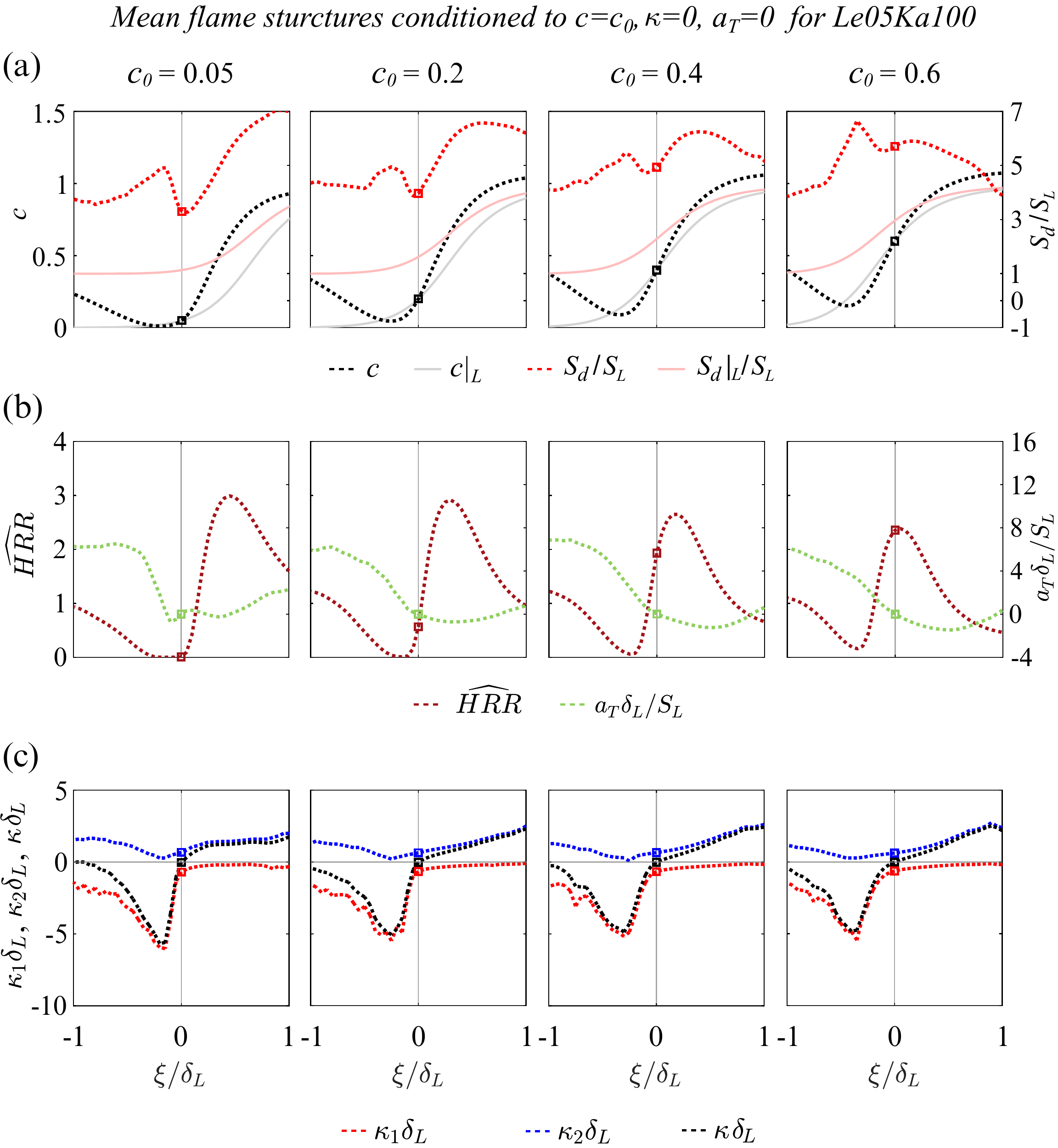}
\vspace{20 pt}
\caption{Mean flame structures of (a) $c$ and $S_d/S_L$ (b) normalized heat release rate, $\widehat{HRR}$ and non-dimensional tangential strain rate, $a_T \delta_L/ S_L$ (c) non-dimensional curvature $\kappa\delta_L$ , non-dimensional minimum and maximum principal curvatures, $\kappa_1\delta_L$ and $\kappa_2\delta_L$ conditioned on $\kappa=0$ and $a_T=0$ at $c=c_0$ for Le05Ka100. The faint red and gray curves in the background in (a) represent the corresponding standard laminar flame structures.}
\label{Fig_16}
\end{figure*}

\begin{figure*}[h!]
\centering
\includegraphics[trim=0cm 0cm 0cm 0cm,clip,width=390pt]{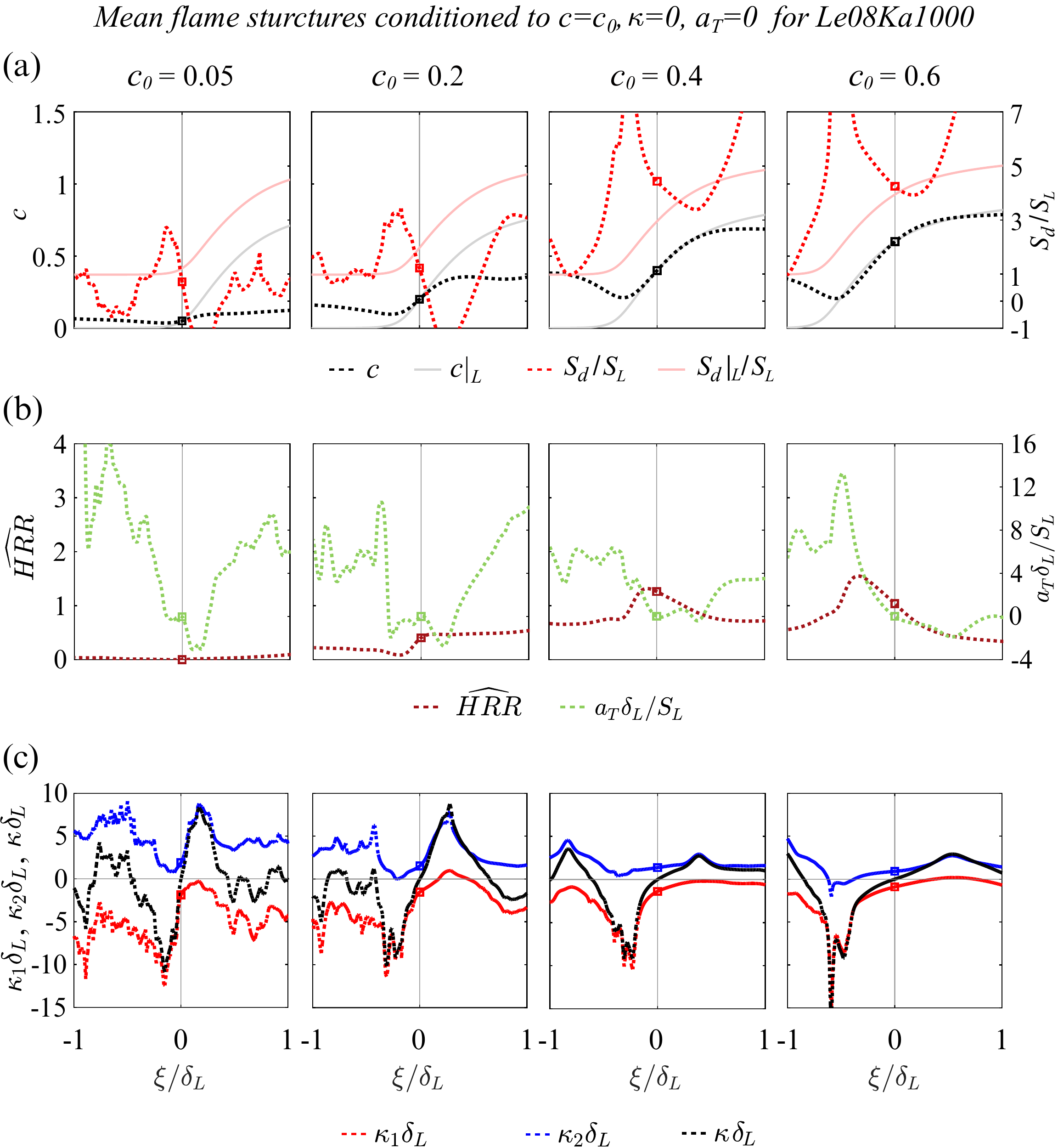}
\vspace{20 pt}
\caption{Mean flame structures of (a) $c$ and $S_d/S_L$ (b) normalized heat release rate, $\widehat{HRR}$ and normalized tangential strain rate, $a_T \delta_L/ S_L$ (c) non-dimensional curvature $\kappa\delta_L$, non-dimensional minimum and maximum principal curvatures, $\kappa_1\delta_L$ and $\kappa_2\delta_L$ conditioned on $\kappa=0$ and $a_T=0$ at $c=c_0$ for Le08Ka1000. The faint red and gray curves in the background in (a) represent the corresponding standard laminar flame structures.}
\label{Fig_17}
\end{figure*}

\begin{figure*}[h!]
\centering
\includegraphics[trim=0cm 0cm 0cm 0cm,clip,width=390pt]{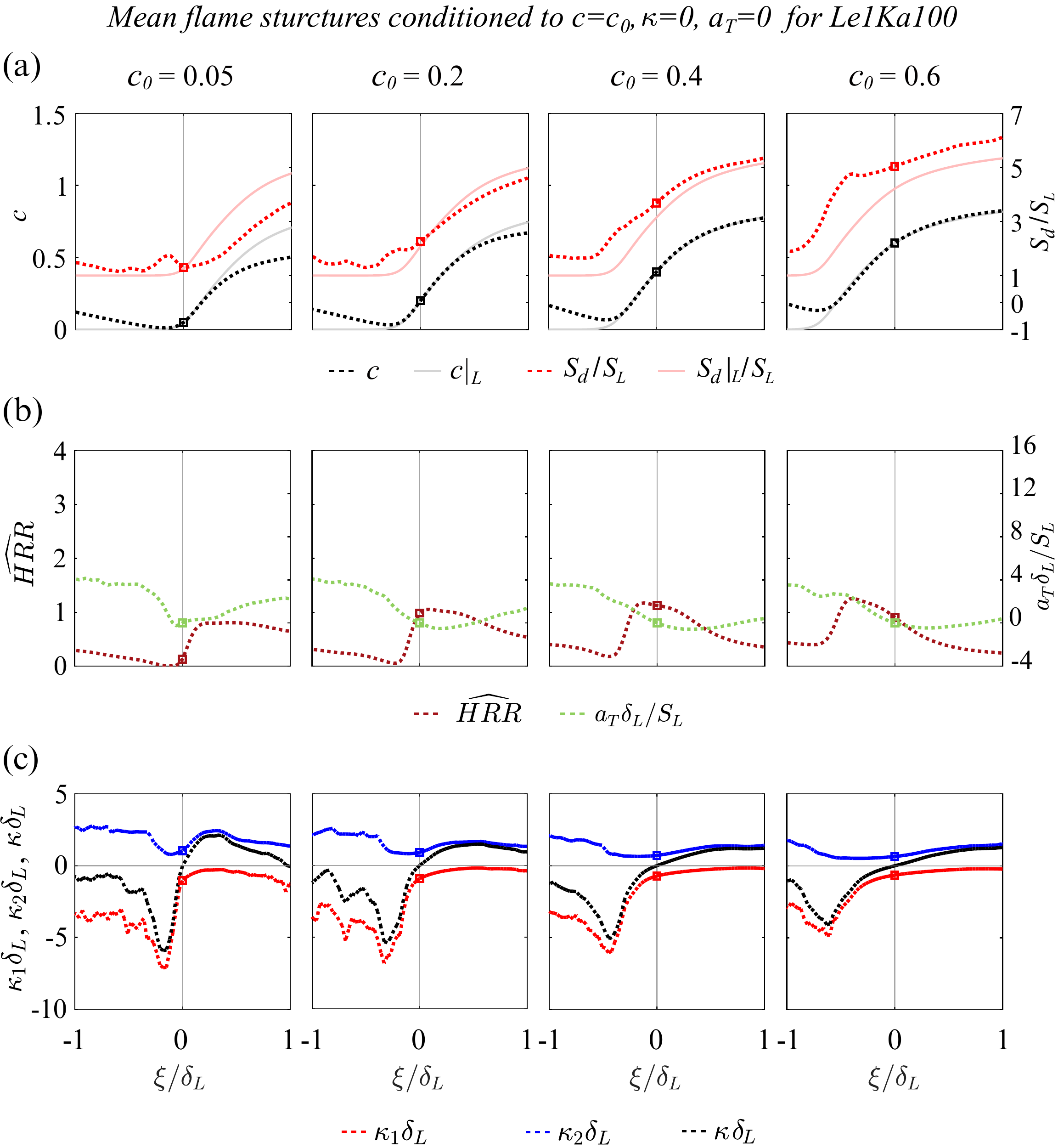}
\vspace{20 pt}
\caption{Mean flame structures of (a) $c$ and $S_d/S_L$ (b) normalized heat release rate, $\widehat{HRR}$ and normalized tangential strain rate, $a_T \delta_L/ S_L$ (c) non-dimensional curvature $\kappa\delta_L$, non-dimensional minimum and maximum principal curvatures, $\kappa_1\delta_L$ and $\kappa_2\delta_L$ in conditioned on $\kappa=0$ and $a_T=0$ at $c=c_0$ for Le1Ka100. The faint red and gray curves in the background in (a) represent the corresponding standard laminar flame structures.}
\label{Fig_18}
\end{figure*}

To isolate the effect of strain rate on increased heat release rate downstream of $\xi/\delta_L=0$ we investigate conditionally averaged flame structures along the local normal directions from the flame surface $c=c_0$ where $\kappa=a_T=0$. The points on the flame surface conditioned on $a_T=0$ are filtered over the range $-0.1<a_T\delta_L/S_L <0.1$ for Le05Ka100 and Le1Ka100 whereas given the intense tangential straining at $Ka\sim\mathcal{O}(1000)$ fewer points are obtained on the flame surface for the given range of $a_T\delta_L/S_L$. To achieve statistical convergence of the structures we increase the range to $-0.5<a_T\delta_L/S_L <0.5$ for Le08Ka1000.  

Figures \ref{Fig_16}, \ref{Fig_17} and \ref{Fig_18} present the triple conditional averaged flame structures for the three DNS cases. In addition to $c=c_0$, and $\kappa=0$, the averaging is further conditioned to $a_T=0$. The rows of these three figures depict the following as a function of $\xi/\delta_L$: (a) $c$, $S_d/S_L$ (b) $\widehat{HRR}$, $a_T\delta_L/S_L$ (c) $\kappa_1\delta_L$, $\kappa_2\delta_L$, $\kappa_2\delta_L$. Columns depict results from each of the isotherms $c_0=0.05,0.2,0.4$ and $0.6$ for the three cases Le05Ka100, Le08Ka1000 and Le1Ka100, respectively. For Le05Ka100 the nature of structures based on $S_d/S_L$, $c$, $\widehat{HRR}$, $\kappa\delta_L$, $\kappa_1\delta_L$ and $\kappa_2\delta_L$ qualitatively resemble those presented in Fig.~\ref{Fig_12}. The $S_d>S_d|_L$ at $\xi/\delta_L=0$ for all $c_0$. The temperature gradient is also higher than the laminar counterpart at $\xi/\delta_L=0$. On the other hand, for $0<\xi/ \delta_L\leq1$, $a_T\delta_L/S_L$ attains very small values, close to zero. In addition, $\kappa\delta_L\approx\kappa_2\delta_L$, i.e., the cylindrical shape of the flame surface with positive curvature, is retained just downstream of $\xi/\delta_L=0$. This shows that the enhanced heat release rate caused by differential diffusion at the positive stretch rate is contributed majorly by positive curvature rather than the tangential strain rate $a_T$. Since the $Le\approx0.5$, the molecular diffusivity of hydrogen exceeds that of the thermal diffusivity of the ultra-lean mixture. Thus, near cylindrical regions with positive curvature ($\kappa \delta_L>1$) allow diffusion of molecular hydrogen from low-temperature zero-curvature regions into the higher temperature regions with positive curvature. This results in enhanced heat released rate and consequently increased temperature and the corresponding gradient. This leads to an enhancement in $\widetilde{S_d}$ at $\xi/\delta_L=0$. Thus the enhancement of $\widetilde{S_{d,0}}$ conditioned on zero tangential strain rate over $S_L$ (black hollow square in Fig.~\ref{Fig_2}) is attributed majorly to differential diffusion at positively curved regions. It should be noted that though the preexisting history effects are not the focus of the present study, we acknowledge their possible contribution to enhancement in $S_d$ as well \cite{im1996response}.

Based on the investigation of the mean flame structures at zero-curvature points \citet{howarth2023thermodiffusively} reported enhanced temperature and heat release rate at zero-curvature regions lying on the mass fraction based iso-scalar surface with $c_0=0.9$. The flame structures were obtained along the paths normal to the flame surface following the gradients of the mass fraction and then averaged to obtain the corresponding mean flame structure \cite{howarth2022empirical}. Given the definition of the constructed path on which the structures were obtained, non-local flame-flame interaction upstream of $\xi/\delta_L=0$ was not captured in their analysis. 
The present finding that enhanced heat release rate due to differential diffusion at large positively curved regions leads to faster propagation of the flame locally ($\xi/\delta_L=0$) is in agreement with \citet{howarth2023thermodiffusively}. However, they argued that after the passage of these positively curved leading points, a wake region with low curvature remains, which is still at superadiabatic temperatures sustaining higher reaction rates than the corresponding laminar case. This results in propagation speeds exceeding $S_L$. Here, we find that in the obtained conditionally averaged structures, it is the positively curved near cylindrical regions and not leading points that are spherical that experience enhanced burning. Moreover, such regions are present downstream of the zero-curvature regions in the conditionally averaged flame structure. The triple conditioned mean flame structures for Le08Ka1000 and Le1Ka100 in Fig.~\ref{Fig_17} and Fig.~\ref{Fig_18} also resemble their double conditioned counterparts presented in Fig.~\ref{Fig_13} and Fig.~\ref{Fig_14} respectively. Near cylindrical positively curved regions are also present with $a_T\delta_L/S_L\approx0$ in the near vicinity of $\xi/\delta_L=0$ in the downstream of it for both cases at all $c_0$. Since the $Le$ is close to unity, a temperature gradient less than ($c_0=0.05$ and $0.2$) or equal ($c_0=0.4$ and $0.6$) to the laminar case at $\xi/\delta_L=0$ is observed for Le08Ka100. For Le1Ka100 at $c_0=0.4$ and $0.6$ occurring downstream of maximum heat release rate, we observe $S_d>S_d|_L$ at $\xi/\delta_L=0$ this is due to the increased contribution from the heat release term to $S_d$ compared to the diffusion term, shown in Eq.~(\ref{Eq_2}).

Finally, it is noteworthy to mention that although planar laminar flames under tangential straining with increased gradients are faster than the standard laminar flame, they do not experience the effect of differential diffusion at positive curvatures. On the other hand, turbulent flame surfaces with zero-curvatures experience enhanced $\widetilde{S_d}$ majorly due to the non-local $Le$ effect rather than the local tangential strain rate. Therefore, the regions with $\kappa=0$ on a turbulent flame surface, on average, can propagate faster than the canonical form under the same tangential strain rates. We observe this for the $c_0$ lying beyond the peak heat release rate, i.e., $c_0=0.6$ for Le08Ka1000 and $c_0=0.4,0.6$ for Le1Ka100 (see Fig.~\ref{Fig_5}b and c). Thus, due to the distinct nature of the mean local flame structure, including the underlying mechanism for $\widetilde{S_d}$ enhancement at zero-curvature regions in turbulent flames, the standalone CFF may not be the best model for predicting $\widetilde{S_{d,0}}$ under the same tangential strain rate condition.

\section{Conclusions} \label{sec: 4}
Localized cylindrical flame-flame interaction at large negative curvatures leads to enhanced flame displacement speeds in near unity Lewis number lean hydrogen-air turbulent flames. Analytical or numerical interaction models can explain such large excursions of $\widetilde{S_d}$. The present study investigates the behavior of $\widetilde{S_d}$ at large negative curvatures in ultra-lean H$_2$-air turbulent flames. Its implications, paired with the Lewis number effects on the conditionally averaged flame structure and consequently the mean local flame displacement speed at zero-curvature, are further explored as well.

To that end, three 3D DNS datasets at different $Le$ and $Ka$ are investigated to understand the $\widetilde{S_d}$ at large negative curvatures as well as at zero-curvatures. Detailed reaction mechanisms are used to model the chemistry for all the datasets. Strong negative flame stretch led to an extremely low heat release rate during flame-flame interaction at large negative curvature. This resulted in a near-perfect agreement of mean $\widetilde{S_d}$ conditioned on curvature with the interaction model, in the asymptotic limit of large negative curvature.  

It is found that most parts of ultra-lean, large $Ka$ hydrogen flames, characterized by $\kappa\approx0$, propagate faster in turbulence when compared to their standard laminar counterpart. However, the canonical configuration of a steady laminar strained flame was insufficient in quantifying and explaining such enhancement. Mapping flame displacement speed (ratios) with the local structure represented by the thermal gradient (ratios), the reason for this difference systematically emerges. All the investigated turbulent flames, irrespective of $Le$ and $Ka$, are characterized by remarkable structures when conditionally averaged from the zero-curvature regions. At zero-curvatures, local broadening persists due to a reversed flame speed gradient resulting from upstream, non-local cylindrical flame-flame interaction. The severity of this effect depends on the $Ka$. This effect is paired with an increased local temperature gradient resulting from differential diffusion at the near cylindrical, positively curved regions downstream of zero-curvature regions. These two effects impart the large $Ka$ turbulent premixed flame its distinct conditionally averaged local structure and the local flame speed. 

This work thus highlights for the first time that mean local flame structure in turbulence results from the aforementioned non-local effects rather than just local tangential straining present in the canonical configurations. The obtained conditionally averaged local flame structure may prove beneficial for modeling thickened flames at high Karlovitz numbers, along with the corresponding local flame displacement speed. This advancement could facilitate more accurate modeling of turbulent flame structures and propagation at both local and global levels and could be instrumental in understanding and mitigating flashbacks and knocking in ultra-lean turbulent combustion of hydrogen.

\section{Acknowledgement}

This research was supported in part by the Natural Sciences and Engineering Research Council of Canada through a Discovery Grant and by King Abdullah University of Science and Technology (KAUST). Computational resources were provided by the SciNet High-Performance Computing Consortium at the University of Toronto and the Digital Research Alliance of Canada (the Alliance).

\appendix




\bibliographystyle{elsarticle-num-names}
\bibliography{sample.bib}







\end{document}